\documentclass[fleqn,usenatbib,useAMS]{mnras}

\usepackage{graphicx}	
\usepackage{amsmath}	
\usepackage{amssymb}	
\usepackage{multicol}        
\usepackage{bm}		
\usepackage{pdflscape}	

\renewcommand{\d}{\mathrm{d}}

\newcommand{\Cl}{C_{\ell}}
\newcommand{\Cltilde}{\tilde{C}_{\ell}}

\newcommand{\Clvec}{\mathbfit{C}_\ell}
\newcommand{\Cltildevec}{\tilde{\mathbfit{C}}_{\ell}}

\newcommand{\yl}{y_{\ell}}

\newcommand{\alm}{a_{\ell m}}

\newcommand{\almE}{a_{\ell m}^{E}}
\newcommand{\almEtilde}{\tilde{a}_{\ell m}^{E}}
\newcommand{\almEprime}{a_{\ell' m'}^{E}}
\newcommand{\almB}{a_{\ell m}^{B}}
\newcommand{\almBtilde}{\tilde{a}_{\ell m}^{B}}
\newcommand{\almBprime}{a_{\ell' m'}^{B}}

\newcommand{\Wllmm}{\, _{\pm 2}W^{\ell \ell'}_{m m'}}
\newcommand{\Mmat}{\mathbfss{M}_{\ell \ell'}}

\newcommand{\ClEE}{C_{\ell}^{EE}}
\newcommand{\ClBB}{C_{\ell}^{BB}}
\newcommand{\ClEB}{C_{\ell}^{EB}}

\newcommand{\Nl}{N_{\ell}}

\newcommand{\Ymat}{\mathbfss{Y}}

\newcommand{\Ylm}{Y_{\ell m}}
\newcommand{\sYlm}{_{s}Y_{\ell m}}
\newcommand{\stwoYlm}{\, _{\pm 2}Y_{\ell m}}
\newcommand{\splustwoYlm}{\, _{+2}Y_{\ell m}}
\newcommand{\sminustwoYlm}{\, _{-2}Y_{\ell m}}
\newcommand{\stwoYlmprime}{\, _{\pm 2}Y_{\ell' m'}}

\newcommand{\nhatvec}{\hat{\mathbfit{n}}}

\newcommand{\Nside}{N_{\mathrm{side}}}
\newcommand{\Npix}{N_{\mathrm{pix}}}
\newcommand{\lmax}{\ell_{\mathrm{max}}}

\newcommand{\fsky}{f_{\mathrm{sky}}}

\newcommand{\Tr}{\mathrm{Tr}}

\newcommand{\sigmaeight}{\sigma_{8}}
\newcommand{\Seight}{S_{8}}

\newcommand{\Omegam}{\Omega_{\mathrm{m}}}

\usepackage{relsize}
\usepackage{xspace}
\newcommand{\Cpp}{C\raisebox{.1ex}{\textbf{++}}\xspace}

\usepackage{fontawesome}

\usepackage{bbold}

\usepackage[T1]{fontenc}
\usepackage{ae,aecompl}

\usepackage{newtxtext,newtxmath}

\title[QML estimators for Stage-IV experiments]{Testing Quadratic Maximum
  Likelihood estimators for forthcoming Stage-IV weak lensing surveys}

\author[A. Maraio et al.]{Alessandro Maraio%
\thanks{Contact e-mail: \href{mailto:maraio@roe.ac.uk}{maraio@roe.ac.uk}},
Alex Hall, Andy Taylor%
\\
Institute for Astronomy, University of Edinburgh, Royal Observatory, Blackford Hill, Edinburgh EH9 3HJ, UK}

\date{Last updated 21 July 2022}

\pubyear{2022}

\begin{document}
\label{firstpage}
\pagerange{\pageref{firstpage}--\pageref{lastpage}}
\maketitle

\begin{abstract}
  Headline constraints on cosmological parameters from current weak lensing
  surveys are derived from two-point statistics that are known to be
  statistically sub-optimal, even in the case of Gaussian fields. We study the
  performance of a new fast implementation of the Quadratic Maximum Likelihood
  (QML) estimator, optimal for Gaussian fields, to test the performance of
  Pseudo-$\Cl$ estimators for upcoming weak lensing surveys and quantify the
  gain from a more optimal method. Through the use of realistic survey
  geometries, noise levels, and power spectra, we find that there is a decrease
  in the errors in the statistics of the recovered $E$-mode spectra to the level
  of $\sim \!\! 20\,\%$ when using the optimal QML estimator over the
  Pseudo-$\Cl$ estimator on the largest angular scales, while we find
  significant decreases in the errors associated with the $B$-modes for the QML
  estimator. This raises the prospects of being able to constrain new physics
  through the enhanced sensitivity of $B$-modes for forthcoming surveys that our
  implementation of the QML estimator provides. We test the QML method with a
  new implementation that uses conjugate-gradient and finite-differences
  differentiation methods resulting in the most efficient implementation of the
  full-sky QML estimator yet, allowing us to process maps at resolutions that are
  prohibitively expensive using existing codes. In addition, we investigate the
  effects of apodisation, $B$-mode purification, and the use of non-Gaussian
  maps on the statistical properties of the estimators. Our QML implementation
  is publicly available and can be accessed from
  \href{https://github.com/AlexMaraio/WeakLensingQML}{\texttt{GitHub}
  \faicon{github}}.
\end{abstract}



\section{An introduction}
\label{sec:Introduction}

Cosmic shear is the study of the coherent distortion in the shapes of background
galaxies due to the matter distribution of the intervening large-scale
structure~\citep{Bartelmann:1999yn,Bartelmann:2010fz,Kilbinger:2014cea}. Since
these distortions are sensitive to the total matter distribution, with
contributions from both ordinary baryonic matter and non-luminous dark matter,
cosmic shear is a powerful probe of dark matter. By measuring the cosmic shear
signal in multiple redshift bins, we can place constraints on the evolution of
structure in the Universe, ultimately placing constraints on the properties of
dark energy - a key goal for cosmology in the current decade.

Given the large quantity of high-precision cosmic shear data that forthcoming
Stage-IV weak gravitational lensing surveys, such as the \textit{Euclid} space
telescope \citep{Euclid:2011zbd}, the Legacy Survey of Space and Time (LSST) at
the \textit{Rubin} observatory \citep{LSSTDarkEnergyScience:2012kar}, and the
\textit{Roman} space telescope \citep{Spergel:2015sza}, are expected to take, it
is important to ensure that we are using the most optimal methods possible
throughout the data analysis pipeline. Given that each of these observatories
will observe and measure the shear of well over one billion galaxies, it is
unfeasible to perform data analysis on each of these individual galaxies. Hence,
some form of data compression steps are needed to make the data processable.
Here, we have investigated the process of compressing maps of the observed
ellipticities of galaxies into two-point summary statistics, namely the power
spectrum. The use of two-point statistics is well motivated because for Gaussian
fields the power spectrum contains all of the information about the field, and
two-point statistics have been extensively studied leading to the development of
robust, well-tested models. The estimation of two-point statistics from data is
an important process as it allows comparisons between observations and values
predicted from cosmological theories to be performed. These comparisons allow us
to constrain the cosmological parameters that feature in the models, and the
cosmic shear power spectrum has been used to obtain competitive
results~\citep{Heymans:2020gsg,DES:2022qpf,Hamana:2019etx}. Recent works have
suggested that there could be a possible tension to the level of $\sim\!3
\sigma$ in the value for the structure of growth parameter, $\Seight \equiv
\sigmaeight \sqrt{\Omegam / 0.3}$, measured between data from cosmic shear
surveys and those from cosmic microwave background experiments
\citep[e.g.][]{Heymans:2020gsg,Abdalla:2022yfr}. Therefore, to help determine if
this tension has physical origins or not, it is essential to ensure that our
analysis methods are as optimal as possible.

The process of compressing maps into two-point summary statistics is crucial,
and so naturally a number of competing methods to do so have been developed and
applied to cosmic shear data. Most notably are the two-point correlation
functions (2PCF) $\xi_{\pm}(\theta)$ \citep{Kaiser:1991qi,Schneider:2002jd}, the
Complete Orthogonal Sets of E-/B-mode Integrals (COSEBIs)
\citep{Schneider:2010pm,KiDS:2020suj}, and the power spectrum coefficients $\Cl$
\citep{Hu:2000ax,Brown:2004jn,Hikage:2010sq}. In this work, we focus on
analysing cosmic shear maps using the power spectrum. The power spectrum has the
advantage that it provides the most direct comparison between theory- and
data-vectors, without the need to perform any additional transformations when
comparing them, as is required for analyses using correlation
functions~\citep{Schneider:2002jd}. Additionally, the power spectrum provides a
cleaner separation between linear and non-linear scales, which aids the
investigation of biases from the non-linear modelling of the matter power
spectrum and intrinsic alignments \citep{DES:2022qpf}, and the scale-dependent
signatures in the power spectrum - for example arising from the properties of
massive neutrinos and baryonic effects. We note that none of the methods
discussed in this work employ the flat-sky approximation, with all quantities
being evaluated on the full, curved-sky.

While power spectrum estimators are a sub-set of two-point correlators, we can
further break down this category of estimators into two main methods: the
Pseudo-$\Cl$ method \citep{Hivon:2001jp,Brown:2004jn}, and the quadratic maximum
likelihood method (QML) \citep{Tegmark:1996qt,Tegmark:2001zv}. In addition,
there is the \texttt{PolSpice} algorithm which uses the correlation functions to
produce estimates of the power spectrum, and is statistically equivalent to the
Pseudo-$\Cl$ method~\citep{Szapudi:2000xj,Chon:2003gx}. The Pseudo-$\Cl$ class of estimators work in
harmonic-space and utilise very efficient spherical harmonic transformation
algorithms which makes this class of estimator extremely numerically efficient
even for high-resolution maps \citep{Hivon:2001jp,Gorski:2004by}. Alternatively,
the QML method works in pixel-space, which results in a much larger
computational demand when compared to Pseudo-$\Cl$ method for the same
resolution maps. Traditionally, this has limited analyses using the QML method
to low-resolution maps only, and thus confined the values for the recovered
power spectrum to low multipoles. Hence, when power spectrum methods have been
applied to existing weak lensing data, the Pseudo-$\Cl$ estimator has been the
method of choice for the vast majority of weak lensing analyses
\citep{HSC:2018mrq,KiDS:2020suj,Nicola:2020lhi,
Garcia-Garcia:2019bku,DES:2022qpf}, primarily using the \texttt{NaMaster} code
which is a fast implementation of the Pseudo-$\Cl$ estimator that can be easily
applied to weak lensing analyses \citep{Alonso:2018jzx}. The Pseudo-$\Cl$
estimator will form part of the analysis pipeline for upcoming weak lensing
surveys~\citep{KiDS:2021opn}. However, it has been shown that while the QML
estimator is optimal, in the sense that it estimates a power spectrum with the
minimal possible errors \citep{Tegmark:1996qt}, it is known that the
Pseudo-$\Cl$ method is not optimal, and thus could be introducing additional
errors into the data.

Another aspect of the data analysis expected to benefit significantly from the
application of an optimal estimator is in $B$-mode measurement. In the limit of
weak gravitational lensing, the produced signal should form a curl-free field,
and thus the predicted $B$-mode signal for cosmic shear should be zero.
Traditionally, a statistically significant detection of $B$-modes would indicate
the presence of unaccounted systematic effects present in the data. However,
with the increased precision of forthcoming Stage-IV experiments, the $B$-mode
signal will be treated as a potential signal that could give hints of new
physical phenomena. An application of using $B$-modes to constrain novel
cosmological models was presented in \cite{Thomas:2016xhb}. Hence, ensuring that
the $B$-mode errors are as small as possible (to help determine if any residual
$B$-mode signal is statistically significant, and to distinguish between
systematic effects and a cosmological $B$-mode signal) is another key feature
for any power spectrum estimation technique that would be applied to upcoming
experimental data. 

Additionally, in a power spectrum analysis a choice for how the survey mask
should be modelled is present. Either the effects of the mask can be deconvolved
from the observed values, giving predictions for the full-sky power
\citep{Hikage:2010sq}, or the effects of the mask can be convolved into the
theory predictions, the so-called `forward-modelling' approach
\citep{KiDS:2021opn}. Here, we focus on the full-sky predictions, which the QML
estimator naturally produces, as this allows for the most straightforward
comparisons between experimental results and theoretical predictions to be made.
This is because there is no need to convolve the theory data-vector with the
mask at every step in an analysis chain when using Monte Carlo Markov Chain
methods, and thus reducing the per-step computational requirements resulting in
faster run-times. We also note that by producing estimates of the full-sky power
spectra, our covariance matrix is less affected by the mask on large scales.

Previous attempts at applying QML methods to existing weak lensing surveys have
found little differences in results when compared to other analysis techniques.
\cite{Kohlinger:2015tza} applied a QML implementation to estimate band powers
from the data from the Canada-France-Hawaii Telescope Lensing Survey (CFHTLenS)
finding general consistency between their QML analysis and all other studies
using CFHTLenS data. This implementation was then applied to data from the first
data release from the Kilo Degree Survey (KiDS-450) in \cite{Kohlinger:2017sxk}.
Here, they again found broadly consistent results between their analysis and
previous works, though finding a slightly smaller value for $\Seight$ which
could be explained by their work using a slightly smaller range of $\ell$
multipole values \citep{vanUitert:2017ieu}. Quadratic and Pseudo-$\Cl$
estimators were applied to cosmic shear measurements performed by the Sloan
Digital Sky Survey in \cite{SDSS:2011gwu}, again finding strong consistency
between the two methods. This demonstrates that analysing weak lensing data
using QML methods provides a strong consistency check between different
two-point estimators and ensuring that results are robust to the different
analysis choices that are required for the different methods. While these
previous analyses of weak lensing data using QML methods have shown strong
consistencies in the main cosmological results, though their application as a
cross-check remains an important use case, we note that the CFHTLenS and
KiDS-450 surveys covered about $154 \, \deg^2$ and $450 \, \deg^2$ of sky,
respectively. These sky areas are around two orders of magnitude smaller than
the expected sky area that forthcoming Stage-IV experiments are expected to
cover. While it has been shown that the QML and Pseudo-$\Cl$ estimators are
statistically equivalent in the high noise
regime~\citep{Efstathiou:2003dj,Efstathiou:2006eb}, the expected noise levels
for forthcoming surveys will be much lower than CFHTLenS and KiDS. Therefore,
the huge increase in statistical precision that forthcoming Stage-IV surveys
will bring means that the use of non-optimal methods (the Pseudo-$\Cl$
estimator) needs to be reassessed and their affects on cosmological constraints
quantified. 

Despite the advantages of quadratic estimators, the development of maximum
likelihood estimators, and in particular their applications to cosmic shear, has
traditionally been less explored than other techniques because of their
computational complexity and associated slowness. In general, they require the
computation and inversion of a dense pixel-space covariance matrix of the map(s)
which is a slow and inefficient process when compared to other analysis methods.
Recent theoretical developments presented in \cite{Horowitz:2018tbe} and
\cite{Seljak:2017rmr}, along with using methods presented in \cite{Oh:1998sr},
have provided a set of key tools that has allowed us to build a new novel QML
implementation that is highly efficient. We note that the construction of the
QML estimator is very analogous to the Wiener filtering of the data, for which
fast implementations have been recently developed and applied to CMB data-sets
\citep{Elsner:2012fe,Bunn:2016lxi,Ramanah:2018enp}. We also note that recent
works have applied quadratic estimators to galaxy clustering in
\cite{Estrada:2021hdo} and \cite{Philcox:2020vbm} applying their quadratic
estimators to data from the VIPERS and BOSS surveys, respectively.

This paper is structured as follows: in
Section~\ref{sec:Power_spectrum_estimators} we present a review of both the QML
and Pseudo-$\Cl$ estimators, including a detailed derivation of the QML method
in Section~\ref{sec:The_qml_estimator}. Then in
Sections~\ref{sec:Conjugate_gradient_method} and~\ref{sec:Finite_diff_fisher} we
present our new highly efficient implementation of the QML estimator.
Section~\ref{sec:Methodology} outlines our methodology for generating mock weak
lensing data, which is used for our results that we present in
Section~\ref{sec:Results}. Our conclusions are presented in
Section~\ref{sec:Conclusions}.

\section{Power spectrum estimators}
\label{sec:Power_spectrum_estimators}

As discussed in Section~\ref{sec:Introduction}, there exists two broad classes
of power spectrum estimation techniques. Here, we first derive the set of key
results for the QML method, and then present a brief review of the Pseudo-$\Cl$
method.

\subsection{The quadratic maximum likelihood estimator}
\label{sec:The_qml_estimator}

Consider a spin-0 input map as a data-vector $\mathbfit{x}$ that has zero mean,
an example of such a map would be convergence estimates in pixels over the sky.
This data-vector length of the number of pixels in the map $\Npix$. We can write
it as
\begin{align}
    \mathbfit{x} = \mathbfit{s} + \mathbfit{n},
\end{align}
where $\mathbfit{s}$ and $\mathbfit{n}$ are the signal and noise data-vectors,
respectively. Assuming that the signal and noise data-vectors are uncorrelated
and both follow the Gaussian distribution, then the likelihood function for the
power spectrum coefficients recovered from the map, $\Cltilde$, is given by
\begin{equation}
    \mathcal{L}(\Cltilde \,|\, \mathbfit{x}) = p(\mathbfit{x} | \Cltilde) = 
    \frac{\exp \left( -\frac{1}{2} \mathbfit{x}^{\dagger} \, \mathbfss{C}^{-1} \, \mathbfit{x} \right)}{\left(2 \pi\right)^{\Npix / 2} \lvert \mathbfss{C} \rvert^{1/2}},
    \label{eqn:Cl_likelihood}
\end{equation}
where $\mathbfss{C}$ is the total pixel-covariance matrix, given as 
\begin{equation}
    \mathbfss{C} = \langle \mathbfit{x} \, \mathbfit{x}^{\dagger} \rangle = \mathbfss{S}(\Cl) + \mathbfss{N},
    \label{eqn:total_cov_mat}
\end{equation}
where $\mathbfss{S}$ is the signal covariance matrix, $\mathbfss{N}$ is the
noise matrix, and $\Cl$ are the fiducial power spectrum coefficients. The signal
covariance matrix can be written as
\begin{equation}
    \mathbfss{S}(\Cl) = \sum_{\ell} \mathbfss{P}_\ell \, \Cl,
\end{equation}
where the $\mathbfss{P}_\ell$ matrices are defined, in pixel-space, as
\begin{equation}
    \mathbfss{P}_\ell \equiv 
    \frac{2\ell + 1}{4 \pi} P_{\ell}(\hat{r}_i \cdot \hat{r}_j),
\end{equation}
where $P_\ell$ are the Legendre polynomials, and $\hat{r}_i$ is the unit vector
for pixel $i$. This matrix can be decomposed into spherical harmonics through
the addition theorem, giving
\begin{equation}
    \mathbfss{P}_{\ell} = \sum_{m=-\ell}^{\ell} \Ylm(\hat{r}_i) \, \Ylm^*(\hat{r}_j).
\end{equation}
We note an important result of
\begin{equation}
    \frac{\partial \mathbfss{C}}{\partial \Cl} = \mathbfss{P}_\ell.
\end{equation}
In harmonic-space, these $\mathbfss{P}_{\ell}$ matrices are simply zeros with
ones along the diagonal corresponding to their $\ell$ value. This makes 
evaluating the signal matrix very easy in spherical-harmonic space.

For uncorrelated noise, the noise matrix $\mathbfss{N}$ in pixel-space is simply
given by the noise variance of the $i$-th pixel along the diagonal with zeros
elsewhere. This makes evaluating the noise matrix very easy in pixel-space. We
note that the QML method may require the manual insertion of a small level of
white noise into the covariance matrix to ensure that it is invertible, as in
some extreme cases the covariance matrix can be
singular~\citep{Bilbao-Ahedo:2017uuk}.

A minimum-variance quadratic estimator of the power spectrum can be
formed as~\citep{Tegmark:1996qt}
\begin{equation}
    y_{\ell} \equiv s_{\ell} - b_{\ell} = \mathbfit{x}^{\mathrm{T}} \, \mathbfss{E}_{\ell} \, \mathbfit{x} - b_{\ell},
    \label{eqn:y_ell_definition}
\end{equation}
where the $\mathbfss{E}_{\ell}$ matrices are given by
\begin{equation}
    \mathbfss{E}_{\ell} = \frac{1}{2} \mathbfss{C}^{-1} \frac{\partial \mathbfss{C}}{\partial \Cl} \mathbfss{C}^{-1} =
    \frac{1}{2} \mathbfss{C}^{-1} \mathbfss{P}_{\ell} \mathbfss{C}^{-1},
\end{equation}
and the noise bias terms $b_{\ell}$ are given as
\begin{equation}
    b_{\ell} = \Tr \left[ \mathbfss{N} \, \mathbfss{E}_{\ell} \right].
\end{equation}
Arranging our $\yl$ and $\Cl$ values into vectors $\mathbfit{y}$ and $\mathbfit{C}$, 
respectively, we can relate our quadratic estimator to the true power spectrum
as
\begin{equation}
    \langle \mathbfit{y} \rangle = \mathbfss{F} \, \mathbfit{C},
    \label{eqn:y_ell_avg}
\end{equation}
where $\mathbfss{F}$ is the Fisher matrix. Formally, this is defined through
the likelihood as~\citep{Bond:1998zw}
\begin{equation}
    \mathbfss{F}_{\ell_1 \ell_2} = - \left\langle \frac{\partial^2 \ln \mathcal{L}}{\partial C_{\ell_1} \, \partial C_{\ell_2}} \right\rangle.
    \label{eqn:Fisher_deriv_likelihood}
\end{equation}
When applying the likelihood of Equation~\ref{eqn:Cl_likelihood}, we find the
Fisher matrix can be written as
\begin{align}
    \mathbfss{F}_{\ell \ell'} &= \frac{1}{2} \Tr \left[ \mathbfss{C}^{-1} \frac{\partial \mathbfss{C}}{\partial C_{\ell}} \mathbfss{C}^{-1} \frac{\partial \mathbfss{C}}{\partial C_{\ell'}} \ \right] \! ,
    \label{eqn:AnalyticClFisher}
    \\
    &= \frac{1}{2} \Tr \left[ \mathbfss{C}^{-1} \, \mathbfss{P}_{\ell} \, \mathbfss{C}^{-1} \, \mathbfss{P}_{\ell'} \right] \!\! . \nonumber
\end{align}
Assuming that $\mathbfss{F}$ is regular, and thus can be inverted, we can form
an estimator for the recovered power spectrum from our map,
$\tilde{\mathbfit{C}}$, as
\begin{equation}
    \tilde{\mathbfit{C}} = \mathbfss{F}^{-1} \, \mathbfit{y}.
\end{equation}
This estimator is unbiased in the sense that its average is the true, underlying
spectrum \citep{Tegmark:1996qt},
\begin{equation}
    \langle \tilde{\mathbfit{C}} \rangle =  \mathbfit{C},
\end{equation}
and it is optimal in the sense that its covariance matrix of our estimator is 
the inverse Fisher matrix $\mathbfss{F}^{-1}$,
\begin{equation}
    \langle (\tilde{\mathbfit{C}} - \mathbfit{C}) (\tilde{\mathbfit{C}} - \mathbfit{C})^{\mathrm{T}} \rangle 
    = \mathbfss{F}^{-1}
\end{equation}
and thus satisfies the Cram\'{e}r-Rao inequality~\citep{Tegmark:1996bz}.

\subsubsection{Extension to spin-2 fields}

Above, we have derived a set of key-results of the QML estimator applied to a
scalar spin-0 field. These set of equations can be extended to cover spin-2
fields, as presented in \cite{Tegmark:2001zv}. Such spin-2 field is cosmic
shear, of which the observed field can be decomposed into two components through
$\bm{\gamma}(\nhatvec) = \gamma_1 (\nhatvec) + i \gamma_2 (\nhatvec)$. The
data-vector will now have length $2 \Npix$, where it will be given by
$\mathbfit{x} = \left\{ \vec{\gamma}, \vec{\gamma}^{*}  \right\}$, and the
covariance matrix (and other associated pixel-space matrices) have dimensions $2
\Npix \times 2 \Npix$, where their block structure will be given by
\begin{equation}
    \mathbfss{C} = \begin{pmatrix}
        \langle \vec{\gamma} \, \vec{\gamma}^{\dagger} \rangle & \langle \vec{\gamma} \, \vec{\gamma}^{\mathrm{T}} \rangle \\
        \langle \vec{\gamma}^{*} \, \vec{\gamma}^{\dagger} \rangle & \langle \vec{\gamma}^{*} \, \vec{\gamma}^{\mathrm{T}} \rangle
    \end{pmatrix}.
\end{equation}
Similarly, the Legendre polynomial matrix $\mathbfss{P}_{\ell}$ will have a
structure for the spin-2 case of~\citep{Tegmark:2001zv}
\begin{equation}
    \mathbfss{P}_\ell = \begin{pmatrix}
        \sum_{m} \splustwoYlm(\hat{r}_i) \, \splustwoYlm^*(\hat{r}_j) &
        \sum_{m} \splustwoYlm(\hat{r}_i) \, \sminustwoYlm^*(\hat{r}_j) \\
        \sum_{m} \splustwoYlm^*(\hat{r}_i) \, \sminustwoYlm(\hat{r}_j) &
        \sum_{m} \splustwoYlm^*(\hat{r}_i) \, \splustwoYlm(\hat{r}_j) \\
    \end{pmatrix},
\end{equation}
and the signal covariance matrix is given by
\begin{equation}
    \mathbfss{S} = \begin{pmatrix}
        \sum_{\ell} \left[\ClEE + \ClBB\right] \mathbfss{P}_{\ell}^{(1, \, 1)} &
        \sum_{\ell} \left[\ClEE - \ClBB\right] \mathbfss{P}_{\ell}^{(1, \, 2)} \vspace*{0.15cm} \\
        \sum_{\ell} \left[\ClEE - \ClBB\right] \mathbfss{P}_{\ell}^{(2, \, 1)} &
        \sum_{\ell} \left[\ClEE + \ClBB\right] \mathbfss{P}_{\ell}^{(2, \, 2)} \\
    \end{pmatrix},
\end{equation}
where $\sYlm$ are the spin-weighted spherical harmonics.

The observed spin-2 shear field can be decomposed on the full-sky into a
curl-free $E$-mode and divergence-free $B$-mode fields through
\citep{Brown:2004jn}
\begin{equation}
    (\gamma_1 \pm i \gamma_2)(\nhatvec) = \sum_{\ell} \sum_{m \,=\, - \ell}^{\ell} 
    \left[ \almE \pm i \almB \right] \stwoYlm(\nhatvec).
    \label{eqn:alm2map}
\end{equation}
This relation can be inverted to give the $\alm$ coefficients on the full-sky as
\begin{equation}
    \almE \pm i \almB = \int \!\! \d \Omega \, \left( \gamma_1 \pm i \gamma_2 \right)\!(\nhatvec)
    \stwoYlm^{*}(\nhatvec).
    \label{eqn:map2alm}
\end{equation}
These $\alm$ coefficients can then be combined to form values for the all-sky
power spectrum through
\begin{equation}
    \Cl^{XY} = \frac{1}{2 \ell + 1} \sum_{m = - \ell}^{\ell} 
    \alm^X \left[\alm^{Y}\right]^{*},
    \label{eq:alm2cl}
\end{equation}
where $X, Y$ denote either $E$ or $B$.

\subsubsection{Affect of the fiducial cosmology}

We note that to construct the covariance matrix, we have to provide our
estimator with a set of fiducial $\Cl$ values. Given that the whole point of the
estimator is to estimate the $\Cl$ values from map(s), of which their underlying
power spectrum are unknown prior to the analysis, it may appear that the
estimated power spectrum will somehow depend on the input cosmology. However,
provided that the same fiducial power spectrum is applied consistently to the
estimator, it will still produce unbiased estimates, but ones that may not
necessarily be truly optimal. An iterative scheme where the results of the
estimator are fed back into the construction of the covariance matrix, with this
process repeating for a number of times, was investigated in
\cite{Bilbao-Ahedo:2021jhn}.

\subsection{Inverting the pixel covariance matrix}
\label{sec:Conjugate_gradient_method}

To evaluate our quadratic estimator, we need an efficient way to evaluate the
set of $y_{\ell}$ values for a given map. These in turn require efficient
evaluation of the inverse-covariance weighted map, $\mathbfss{C}^{-1}
\mathbfit{x}$. Na\"ively, one may want to compute these terms through first
evaluating the total covariance matrix $\mathbfss{C}$ and then inverting it.
However, as we have seen through Equation~\ref{eqn:total_cov_mat},
$\mathbfss{C}$ is made up of both the signal and noise covariance matrices,
resulting in there not being a single efficient basis to evaluate $\mathbfss{C}$
in without having to resort to using expensive massive matrix multiplications
using matrices of spherical harmonics, $\Ymat$. Since these operations scale as
$\mathcal{O}(\Npix^3)$, this is an important limiting factor to the resolution
that can be obtained with traditional QML estimation techniques. 

An alternative approach that negates the need to evaluate the total covariance
matrix is required to obtain competitive resolution results using QML methods.
Previous attempts at this problem have used either Newton-Raphson iteration
techniques to find the root of $\partial \mathcal{L} / \partial \Theta_{\ell} =
0$ \citep{Bond:1998zw,Seljak:1997ep,Hu:2000ax}, or alternatively used conjugate
gradient techniques \citep{Oh:1998sr}, both of which avoid the need to directly
evaluate and invert the covariance matrix and thus offers significantly better
computational performance. Alternative techniques also include an iterative
scheme presented in \cite{Pen:2003mu} and renormalisation-inspired methods
presented in \cite{McDonald:2018mfm,McDonald:2019efe}. Here, we employ the
conjugate-gradient approach, although minimisation approaches have also been
shown to give good results \citep{Horowitz:2018tbe}.

The conjugate gradient method involves an iterative scheme to find the solution
vector $\mathbfit{z}$ that solves the linear equation
\begin{equation}
    \mathbfss{C} \mathbfit{z} = \mathbfit{x}
\end{equation}
for a given covariance matrix and input maps. Since we can split the pixel
covariance matrix into a signal part, which is best suited to harmonic-space,
and a noise part, which is best represented in pixel-space this allows us to
rapidly compute the action of our trial vector $\mathbfit{z}$ on the covariance
matrix through
\begin{equation}
    \mathbfss{S} \mathbfit{z} + \mathbfss{N} \mathbfit{z} = \mathbfit{x}.
\end{equation}
We can rapidly transform our trial vector $\mathbfit{z}$ between pixel- and
harmonic-space through efficient spherical harmonic transform functions
\texttt{map2alm} \& \texttt{alm2map} from the \texttt{HEALPix} 
library~\citep{Gorski:2004by}. Qualitatively, our iterative scheme proceeds
along the following steps:
\begin{enumerate}
    \item Convert our map-based data-vector $\mathbfit{x}$ into a set of $\almE$
    and $\almB$ values through the use of \texttt{map2alm} which implements
    Equation~\ref{eqn:map2alm},
    \item Re-scale the $\alm$ values with the input fiducial power spectrum
    $\ClEE$ and $\ClBB$, respectively,
    \item Generate a new set of spin-2 maps with these new $\alm$ coefficients
    to obtain the contribution from the cosmological signal using
    \texttt{alm2map} which implements Equation~\ref{eqn:alm2map},
    \item Take our original data-vector $\mathbfit{x}$ and multiply all elements
    by the noise variance in the respective pixel to obtain the noise contribution,
    \item Finally sum the signal and noise contributions giving a final set of
    two maps in pixel space.
\end{enumerate} 

We used the \texttt{Eigen}\footnote{\url{https://eigen.tuxfamily.org}} \Cpp
linear algebra package to perform our conjugate gradient computations resulting
in a quick and efficient numerical implementation.

Since we are now only computing the action of the covariance matrix on our trial
vector, instead of explicitly computing the full form of the covariance matrix,
we find that our method provides much better scaling to higher map resolutions
than previous implementations. We explore the speed and memory performance of
our new estimator in Section~\ref{sec:QML_estimtors_benchmark}. In our analysis,
we used map resolutions of $\Nside = 256$, which compares to a maximum of
$\Nside = 64$ that was explored in previous QML
implementations~\citep{Bilbao-Ahedo:2021jhn}.

In general, the conjugate gradient technique can benefit greatly from an
appropriate choice of matrix preconditioner \citep{Oh:1998sr}. Given a linear
system $\mathbfss{A} \mathbfit{x} = \mathbfit{b}$, the preconditioner matrix
$\tilde{\mathbfss{A}}$ should be such that $\tilde{\mathbfss{A}}^{-1}
\mathbfss{A} = \mathbfss{I} + \mathbfss{R}$, where $\mathbfss{I}$ is the
identity matrix and $\mathbfss{R}$ is a matrix whose eigenvalues are all less
than unity. This minimises the number of iterations required for the conjugate
gradient method to converge, and thus can offer significant performance
improvements if properly set. Since our method requires a strictly diagonal
preconditioner, this placed strict constraints on the form and values of the
preconditioner. We investigated the use of different values for the diagonal of
the preconditioner finding little change in the performance of the iterative
method. Thus we used the identity matrix as our preconditioner.

\subsection{Forming the Fisher matrix}
\label{sec:Finite_diff_fisher}

Since the covariance matrix of our quadratic estimator is the inverse Fisher
matrix, we can get estimates for the estimator's errors through computation of
this Fisher matrix. Direct computation of the Fisher matrix through
Equation~\ref{eqn:AnalyticClFisher} requires many massive $2 \Npix \times 2
\Npix$ matrix multiplications and inversions, which has
$\mathcal{O}(\Npix^{3})$ scaling, even for efficient implementations of this
technique~\citep{Bilbao-Ahedo:2021jhn}. Thus, this direct-evaluation technique
becomes unfeasible for map resolutions above about $\Nside = 64$ for Stage-IV
cosmic shear experiments. Hence, to get power spectrum estimates for
higher-resolution maps, which increases the range of $\ell$-values that we can
estimate the power spectrum over, an alternative method to direct computation is
needed. 

We note that the Fisher matrix is related to the second-order derivative of the
likelihood function through Equation~\ref{eqn:Fisher_deriv_likelihood}. Taking a
single derivative of the likelihood yields
\begin{equation}
    \frac{\partial \ln \mathcal{L}}{\partial \Theta_\ell} = s_\ell - b_\ell - \Tr\!\left[\mathbfss{S} \, \mathbfss{E}_{\ell}\right]
    \label{eqn:Likelihood_one_diff}
\end{equation}
where $s_{\ell}$ is our quadratic form of the map as introduced in
Equation~\ref{eqn:y_ell_definition}.
Therefore, to evaluate our Fisher matrix, we wish to take a further derivative
of the above quantity. To do so, we can use the method of finite-differences as
shown in \cite{Seljak:2017rmr}, which yields
\begin{align}
    \left\langle \frac{\partial^2 \ln \mathcal{L}}{\partial \Theta_{\ell} \, \partial \Theta_{\ell'}} \right\rangle &= 
    - \Tr \! \left[\mathbfss{C}^{-1} \, \mathbfss{P}_{\ell} \, \mathbfss{C}^{-1} \, \mathbfss{P}_{\ell'} \right] +
    \frac{1}{2} \Tr \! \left[\mathbfss{C}^{-1} \, \mathbfss{P}_{\ell} \, \mathbfss{C}^{-1} \, \mathbfss{P}_{\ell'}\right]\!\!, \\
    &= -\frac{1}{2} \Tr \left[\mathbfss{C}^{-1} \, \mathbfss{P}_{\ell} \, \mathbfss{C}^{-1} \, \mathbfss{P}_{\ell'} \right]\!,
\end{align}
where the first trace term comes from the differentiation of the quadratic term
$s_{\ell}$ and the second trace arises from the differentiation of the other two
terms in Equation~\ref{eqn:Likelihood_one_diff}. Since we note that the
differentiation of just the $s_{\ell}$ term alone yields twice the negative
Fisher matrix, we can form an estimator for the Fisher matrix using just this
term. Therefore, we can use the method of finite differences to differentiate
$s_{\ell}$ to give~\citep{Seljak:2017rmr,Horowitz:2018tbe}
\begin{equation}
    F_{\ell \ell'} \, \Delta \Theta_{\ell'} = -\frac{1}{2} \left[
    \langle s_{\ell}(\Theta_{\mathrm{fid}} + \Delta \Theta_{\ell'}) \rangle 
    - \langle s_{\ell}(\Theta_{\mathrm{fid}}) \rangle \right].
    \label{eqn:FisherMatrixEstimation}
\end{equation}
Here, we are manually injecting power into a specific $\ell$-mode (given as
$\Delta \Theta_{\ell'}$), generating a map with the modified power spectrum, and
recovering the set of $s_{\ell}$ values. This gives the estimate of our Fisher
matrix associated where we are averaging over many realisations of maps
generated with the specified power spectrum, and $\Theta_{\mathrm{fid}}$ is our
original best-guess for the power spectrum coefficients used when building the
covariance matrix $\mathbfss{C}$.

This approach of using finite-differences to estimate the Fisher matrix performs
much faster than the brute-force calculation, as described in
Equation~\ref{eqn:AnalyticClFisher}, due to our ability to estimate the Fisher
matrix directly from the $s_{\ell}$ values, which are vector quantities and for
which we already have an efficient method to compute though the
conjugate-gradient method, and so we avoid having to compute the matrices and
matrix products of Equation~\ref{eqn:AnalyticClFisher}.

Here, the amount of power injected into the maps at the specific $\ell$-mode is
a free parameter of the method. We used a value of $\Delta \Theta_{\ell} =
10^{7} \, \Theta^{\mathrm{fid}}_{\ell}$ and verified that our results were
insensitive to the choice of this value, provided that it is sufficiently large.

Note that in our analysis presented in this paper, we are not able to recover
any of the covariances associated with any of the $EB$ modes. This is because
these modes are not linearly independent of either the $EE$ or $BB$ spectra, and
thus with our choice of fiducial spectrum containing zero $B$-modes we cannot
inject power into the $EB$ modes. $EB$-spectra can be obtained by setting the
fiducial $B$-mode power to small non-zero values, for example
\cite{Horowitz:2018tbe} use a $B$-mode spectra that has the same shape as their
$E$-mode spectra but has an amplitude that is $10^{-5}$ times smaller. Since we
used zero $B$-mode power as our fiducial model, we are unable to report on any
$EB$ results in this work.

Our new code implementing these approaches is publicly available and can be
downloaded from \texttt{GitHub}:
\href{https://github.com/AlexMaraio/WeakLensingQML}{\texttt{https://github.com/AlexMaraio/WeakLensingQML}
\faicon{github}}.

\subsection[Review of the PseudoCl estimator]{Review of the Pseudo-$\Cl$ estimator}
\label{sec:Review_of_pseudocl}

We refer the reader to \cite{Alonso:2018jzx,Leistedt:2013gfa}, and references
therein, for detailed reviews of the Pseudo-$\Cl$ method, but here we discuss the
key features of the estimator.

Since we cannot observe the shear field on the full-sky, our observed field is
modulated through some window function, $\mathcal{W}(\nhatvec)$, through
$\tilde{\bm{\gamma}}(\nhatvec) = \mathcal{W}(\nhatvec) \bm{\gamma} (\nhatvec)$.
Throughout this work, we consider binary masks only, and thus
$\mathcal{W}(\nhatvec)$ is either zero or one. This turns the recovered harmonic
modes into `pseudo modes', given as  
\begin{equation}
    \almEtilde \pm i \almBtilde = \int \!\! \d \Omega \,\, 
    \mathcal{W}(\nhatvec) \left( \gamma_1 \pm i \gamma_2 \right)\!(\nhatvec)
    \stwoYlm^{*}(\nhatvec),
\end{equation}
where we are denoting quantities evaluated on the cut-sky with a tilde. These
pseudo-modes are related to the underlying modes through
\begin{equation}
    \almEtilde \pm i \almBtilde = \sum_{\ell', \, m'} (\almEprime \pm i \almBprime) 
    \Wllmm,
\end{equation}
where $\Wllmm$ is the convolution kernel for our window function, which can be
written as \citep{Brown:2004jn}
\begin{align}
    \Wllmm &= \int \!\! \d \Omega \,
    \stwoYlmprime(\nhatvec) \, \mathcal{W}(\nhatvec) \, \stwoYlm^{*}(\nhatvec).
\end{align}
Expanding the window function in spherical harmonics and evaluating the
integrals yields
\begin{equation}
    \begin{aligned}
        \Wllmm = \sum_{\ell'', \, m''} (-1)^{m}
        \sqrt{\frac{(2\ell + 1)(2 \ell' + 1)(2 \ell^{''} + 1)}{4 \pi}} \,
        \mathcal{W}_{\ell'' \! m''} \\
        \times 
        \begin{pmatrix}
            \ell & \ell' & \ell'' \\
            \pm 2 & \mp 2 & 0
        \end{pmatrix} 
        \begin{pmatrix}
            \ell & \ell' & \ell'' \\
            \pm m & \mp m' & m''
        \end{pmatrix},
    \end{aligned}
\label{eqn:pcl_mixing_matrix}
\end{equation}
where $\mathcal{W}_{\ell'' m''}$ is the spin-0 spherical harmonic
transform of the mask and the terms in brackets are the Wigner-$3j$ symbols.

When forming the pseudo-multipoles, there is a freedom to add pixel weights
through the window function, for example using inverse-variance weighting
scheme, though this is typically not done in large-scale structure applications.

Combining the three shear spectra into a single vector, $\Clvec = \left( \ClEE,
\, \ClEB, \, \ClBB \right)$, we find that the cut-sky power spectrum can be
written in terms of the full-sky power spectrum through
\citep{Brown:2004jn,Hikage:2010sq}
\begin{equation}
    \Cltildevec = \sum_{\ell'} \Mmat \, \mathbfit{C}_{\ell'},
    \label{eqn:PCl_cutskycls}
\end{equation}
where $\Mmat$ is the mode coupling matrix. Provided that $\Mmat$ is invertible,
which is only the case when there is enough sky area such that the two-point
correlation function $C(\theta)$ can be evaluated on all angular scales
\citep{Mortlock:2000zw}, we can invert this relationship to give an estimate
of the full-sky spectra from the pseudo-modes,
\begin{equation}
    \Clvec = \sum_{\ell'} \Mmat^{-1} \, \tilde{\mathbfit{C}}_{\ell'}.
    \label{eqn:PCl_allskycls}
\end{equation}
This is the final expression for the estimated power spectrum of a map using the
Pseudo-$\Cl$ method. An alternative strategy that avoids this inversion is the
forward-modelling of the mask's mode-coupling matrix into the theory power
spectrum values though Equation~\ref{eqn:PCl_cutskycls}.

Since our analysis was focused on the \textit{errors} associated with the
recovered power spectrum, a detailed description of the covariance matrix
associated with the Pseudo-$\Cl$ estimator is worthy of discussion. 
In general, the exact analytic covariance of two Pseudo-$\Cl$ fields involve
terms of the form \citep{Brown:2004jn,Euclid:2021ilj}
\begin{multline}
    \mathrm{Cov}\! \left[\tilde{C}_{\ell}, \, \tilde{C}_{\ell'} \right] \sim \\
    \sum_{\ell_1, \, \ell_2} C_{\ell_1} C_{\ell_2}
    \sum_{\substack{m, \, m', \\ m_1, \, m_2}} 
    W^{\ell \ell_1}_{m m_1} \, 
    \left( W^{\ell' \ell_1}_{m' m_1} \right)^{*} \, 
    W^{\ell' \ell_2}_{m' m_2} \, 
    \left( W^{\ell \ell_2}_{m m_2} \right)^{*}
    \label{eqn:exact_pcl_covariance}
\end{multline}
Computing this involves summing $\mathcal{O}(\lmax^{6})$ terms which becomes
computationally intractable for even moderate-resolution maps. Hence, certain
assumptions are used to speed up this calculation. The principle of these is the
narrow-kernel approximation, which assumes that the power spectrum of the mask
has support only over a narrow range of multipoles when compared to the power
spectrum \citep{Efstathiou:2003dj,Garcia-Garcia:2019bku}. This involves making
substitutions of the form $\left\{ C_{\ell_1}, \, C_{\ell_2} \right\}
\rightarrow \left\{ C_{\ell}, \, C_{\ell'} \right\}$, and so the power spectrum
terms in Equation~\ref{eqn:exact_pcl_covariance} can be extracted, and then the
symmetric properties of the convolution kernels can be used to simplify the
summations. This approximation is known to be inaccurate on large scales, though
it has been shown that this has negligible impact on parameter constraints, and
for power spectra that contain $B$-modes \citep{Garcia-Garcia:2019bku}.
Alternatively, Gaussian covariances can be estimated from an ensemble of
realisations. While this produces a more accurate estimate of the covariance
matrix, especially for low multipoles, it is much more computationally demanding
due to the large number of realisations required in the ensemble - especially
for accurately determining the off-diagonal elements of the covariance matrix.

\section{Methodology}
\label{sec:Methodology}

Our aim is to investigate to what extent that QML estimators can give improved
statistical errors on the recovered shear power spectra compared to Pseudo-$\Cl$
methods. We will test the estimators on a set of mock shear maps. In this
section, we describe the fiducial setup of these mocks.

\subsection{Theory power spectrum}
\label{sec:Theory_power_spectrum}

For our analysis, we used a single redshift bin with sources following a 
Gaussian distribution centred at $z=1$ with standard deviation of 
$\sigma_z = 0.15$. We used fiducial cosmological values of 
$h = 0.7$, $\Omega_\mathrm{c} = 0.27$,
$\Omega_\textrm{b} = 0.045$, $\sigma_{8} = 0.75$, $n_{s} = 0.96$, and massless
neutrinos.

\begin{figure}
    \includegraphics[width=\columnwidth]{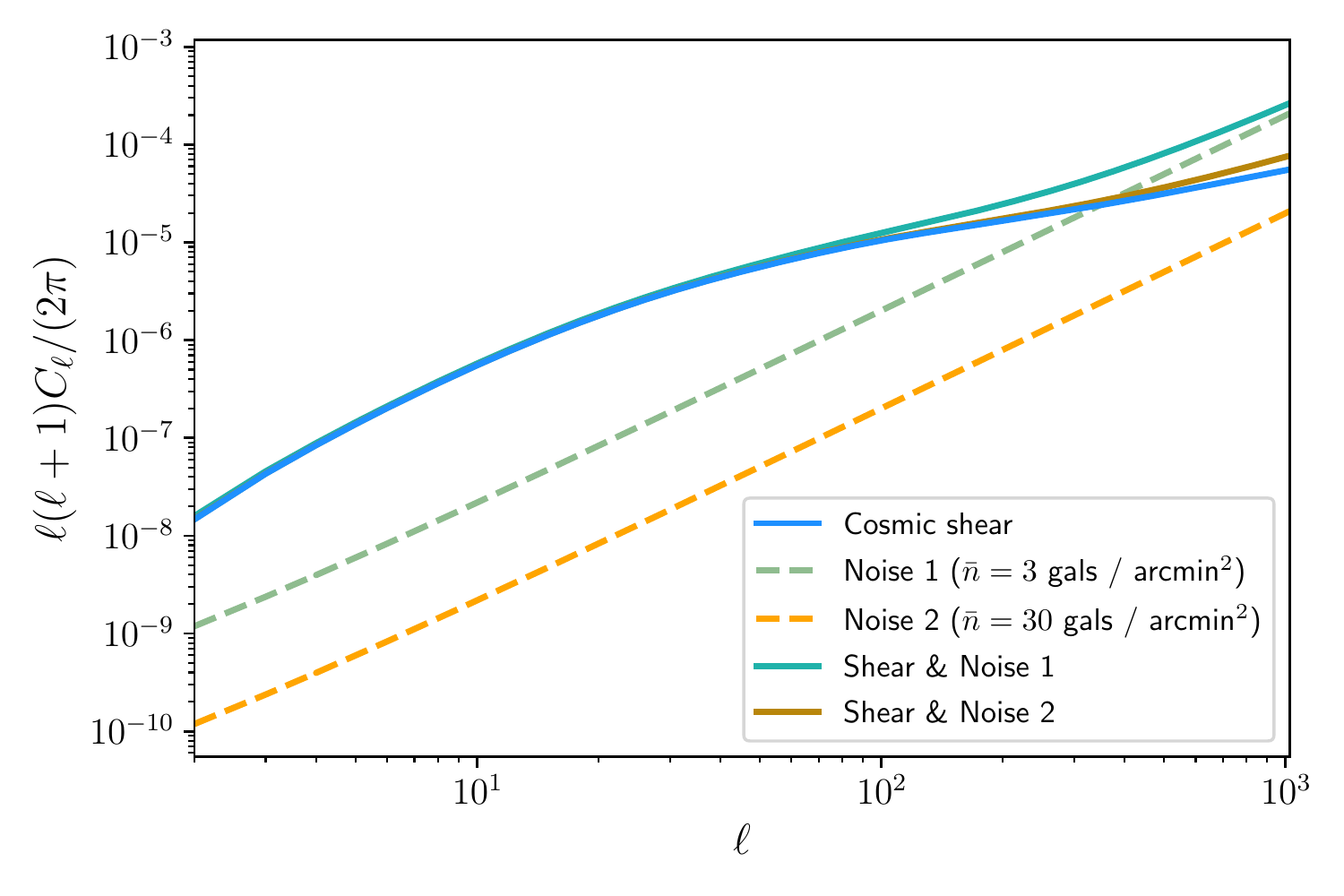}
    \vspace*{-0.75cm}
    \caption{Plot of the fiducial power spectrum values for the cosmic shear
      signal for our single bin of source galaxies (blue curve) that we model as
      following a Gaussian distribution centred at $z=1$ and width $\sigma_z =
      0.15$. We also plot the power spectrum of the shape noise corresponding to
      number densities of $\bar{n} = 3\,\textrm{gals / arcmin}^{2}$ (dashed
      green curve) and $\bar{n} = 30\,\textrm{gals / arcmin}^{2}$ (dashed orange
      curve), and the combined signal and noise spectra (solid green and orange
      curves).}
    \label{fig:FiducialPowerSpectrum}
\end{figure}

The cosmic shear theory signal for this distribution of sources was calculated
using the Core Cosmology Library (\texttt{CCL})
\citep{LSSTDarkEnergyScience:2018yem}. This implements the standard prescription
for the weak lensing power
spectrum~\citep{Bartelmann:1999yn,Bartelmann:2010fz,Kilbinger:2014cea}, where
the convergence power spectrum can be written in natural units where $c=1$ as
\begin{equation}
    C^{\kappa \kappa}_{\ell} = \frac{9}{4} \Omegam^2 \, H_{0}^{4} \! \int_{0}^{\chi_{h}} \!\!\!\!
    \d \chi \, \frac{g(\chi)^2}{a^2(\chi)} \,
    P_\delta \!\left(\! k=\frac{\ell}{f_{K}(\chi)}, \, \chi \! \right)\!,
\end{equation}
where $a(\chi)$ is the scale factor, $P_{\delta}$ is the non-linear matter power
spectrum, $f_{K}$ is the comoving angular diameter distance, and $g(\chi)$ is
the lensing kernel given as
\begin{equation}
    g(\chi) = \int_{\chi}^{\chi_{h}} \!\! \d \chi' \, n(\chi') \frac{f_{K}(\chi' - \chi)}{f_{K}(\chi')},
\end{equation}
where $n(\chi)$ is the number density of source galaxies. The convergence power
spectrum can be transformed into values for the $E$-mode power through
\citep{Hu:2000ee}
\begin{equation}
    \ClEE = \frac{(\ell - 1)(\ell + 2)}{\ell (\ell + 1)} C_{\ell}^{\kappa \kappa}.
\end{equation}
A plot of the $\ClEE$ power spectrum used, including the contribution from shape
noise (described below), is shown in Figure~\ref{fig:FiducialPowerSpectrum}.

Shape noise from the intrinsic ellipticity dispersion of galaxies is an
important factor in cosmic shear analyses. We modelled it as a flat
power-spectrum with value $\Nl$ given as 
\begin{equation}
    \Nl = \frac{\sigma_{\epsilon}^2}{\bar{n}},
    \label{eqn:Noise_power_spectrum}
\end{equation}
where $\sigma_{\epsilon}$ is the standard deviation of the intrinsic galaxy
ellipticity dispersion per component, and $\bar{n}$ is the expected number of
observed galaxies per steradian. For our main analysis, we assume
\textit{Euclid}-like values where it is expected that 30 galaxies per square
arcminute will be observed and divided into ten equally-populated photometric
redshift bins, giving $\bar{n} = 3\,\mathrm{gals /
arcmin}^{2}$~\citep{Euclid:2011zbd}. We investigate the effect of not splitting
the sources into different bins, giving rise to a much lower noise level where
$\bar{n} = 30\,\mathrm{gals / arcmin}^{2}$, in
Section~\ref{sec:Varying_noise_levels}. We take $\sigma_{\epsilon} = 0.21$. 

The shape noise spectrum produces a noise matrix with components given by
\begin{equation}
    N_{ij} = \frac{\sigma_{\epsilon}^2}{n_i} \delta_{ij},
\end{equation}
where $i, j$ are pixel indices, and $n_i$ is the expected number of galaxies in
the $i$-th pixel, which we are assuming is constant and related to $\bar{n}$
through the area of each pixel.

Figure~\ref{fig:FiducialPowerSpectrum} shows the contribution to the total
signal from cosmic shear alone and the shape noise. For the case where we
consider an observed source galaxy density of $\bar{n} = 3\,\textrm{gals /
arcmin}^{2}$, we see three distinct regions: the first is for $\ell \lesssim
200$ where the cosmic shear signal dominates, and thus the the uncertainties are
dominated by cosmic variance, the second is an intermediate set of scales where
the cosmic shear and noise have roughly equal amplitude, and the third is for
scales above $\ell \gtrsim 400$ where the noise dominates. Since we consider the
statistics of our estimators up to a maximum multipole of $\lmax = 512$, this
choice of noise level allows us to test the behaviour of our estimators in these
three regions. Hence, we are sensitive to any differences in the statistics that
might arise in the different regimes. For the case where $\bar{n} =
30\,\textrm{gals / arcmin}^{2}$, we see that we are signal-dominated over our
entire multipole range.

\subsection{Survey geometry}

Since much of the comparison between our two power spectrum estimation
techniques will depend on the specific geometry of the sky mask used, we needed
to use a single, generic mask that can be applied consistently to both
estimators to highlight the effects of the estimators only. For our analysis, we
generated a custom mask that would be applicable to a space-based full-sky weak
lensing observatory. This comprises of a main cut that corresponds to the
galactic-plane combined with a slightly narrower cut that corresponds to the
ecliptic-plane. These two features alone capture the majority of features that
are expected for a \textit{Euclid}-like survey and so our simple model for the
mask will yield representative results. 

In a weak lensing analysis, stars that are present in the data need to be masked
out due to their detrimental effects on determining the shapes of the lensed
galaxies. In our analysis, we looked at the effects of our estimators with and
without stars to see if the inclusion of stars makes any meaningful difference
in either the errors or induced mode-coupling of the recovered power spectra.

\begin{figure}
    \includegraphics[width=\columnwidth]{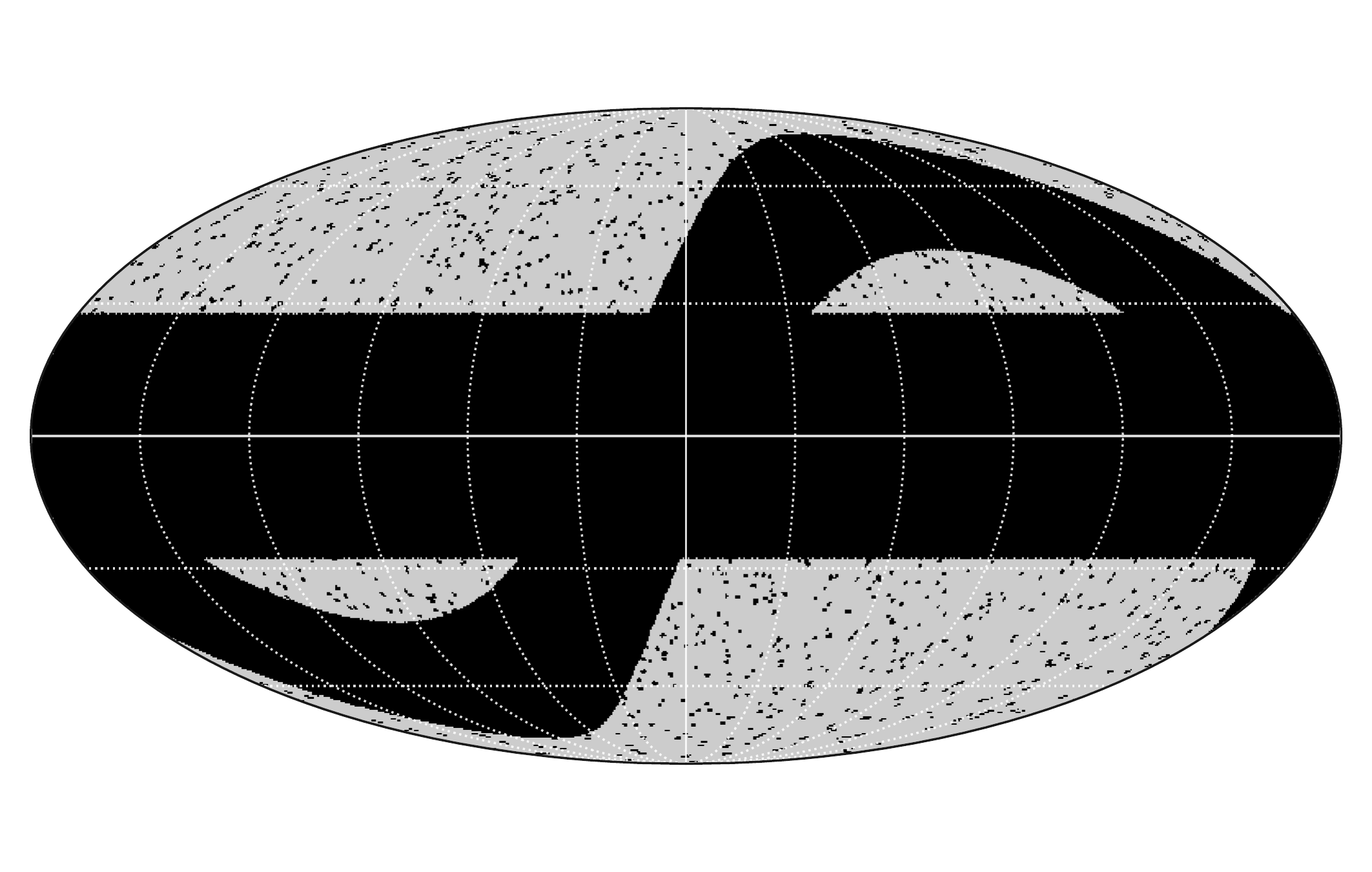}
    \vspace*{-1.0cm}
    \caption{Plot of the sky mask used in our analysis applicable for a
      space-based weak lensing experiment. Here we see the main galactic-cut as
      the thick horizontal band, the ecliptic-cut as the slightly thinner
      sinusoidal band, and our star mask consisting of random circular cut-outs
      that were generated through the prescription of
      Section~\ref{sec:Star_mask_generation}. The galactic- and ecliptic- cuts
      dominate the large-scale behaviour of the mask, whereas the star mask
      introduces strong small-scale effects.}
    \label{fig:SkyMaskWithStars}
\end{figure}

The sky masked used in our analyses is shown in
Figure~\ref{fig:SkyMaskWithStars}.

\subsubsection{Star mask generation}
\label{sec:Star_mask_generation}

We investigated the statistics of our estimators at a map resolution of $\Nside
= 256$. This corresponds to a pixel angular scale of $14\arcmin$. The
prescription described in \cite{Martinet:2020mqm} can be followed to generate a
realistic \textit{Euclid}-like star mask. This involves modelling stars as disks
that are distributed randomly on the sky and that have a radii drawn from a
random uniform distribution taking values between $0.29\arcmin$ and
$8.79\arcmin$. Stars can be placed on a map until the desired sky area covered
by stars is reached. However, we note that this distribution of radii of stars
is smaller than the pixel scale for our map resolution, and so a star mask
generated using such values as presented in \cite{Martinet:2020mqm} would give
rise to under-sampling in the star mask produced, as all stars would be single
pixels, which was found to induce errors in the Pseudo-$\Cl$ estimator.

Since we are not after an exact realistic distribution of stars in our analysis,
we can instead base our star mask on the distribution of `blinding stars', or
avoidance areas, that are expected to be encountered for a space-based
observatory~\citep{Euclid:2021icp}. These avoidance areas are expected to have
an average area of $0.785\,\mathrm{deg}^2$ and total an area of
$635\,\mathrm{deg}^2$ over the expected survey area. Assuming that these
avoidance areas can be modelled as disks, this corresponds to an average radii
of $30\arcmin$, and so is large enough to cover multiple pixels in our limited
resolution maps. This approximate treatment should capture the main features
brought to the analysis by a more realistic star mask.

We used an edited version of the \texttt{GenStarMask} utility as provided with
the \texttt{Flask}\footnote{\url{https://github.com/ucl-cosmoparticles/flask}}
code \citep{Xavier:2016elr} to generate our star mask using the avoidance area
specification. Our edits were made to draw the radii from a uniform distribution
rather than a log-normal. To add some scatter to our avoidance area mask, we
generated the disks with radii between $25\arcmin$ and $35\arcmin$. To match the
desired total avoidance area, avoidance areas were added until they covered
$5\,\%$ of the full-sky.

\subsubsection{Power spectrum of mask}

\begin{figure}
    \includegraphics[width=\columnwidth]{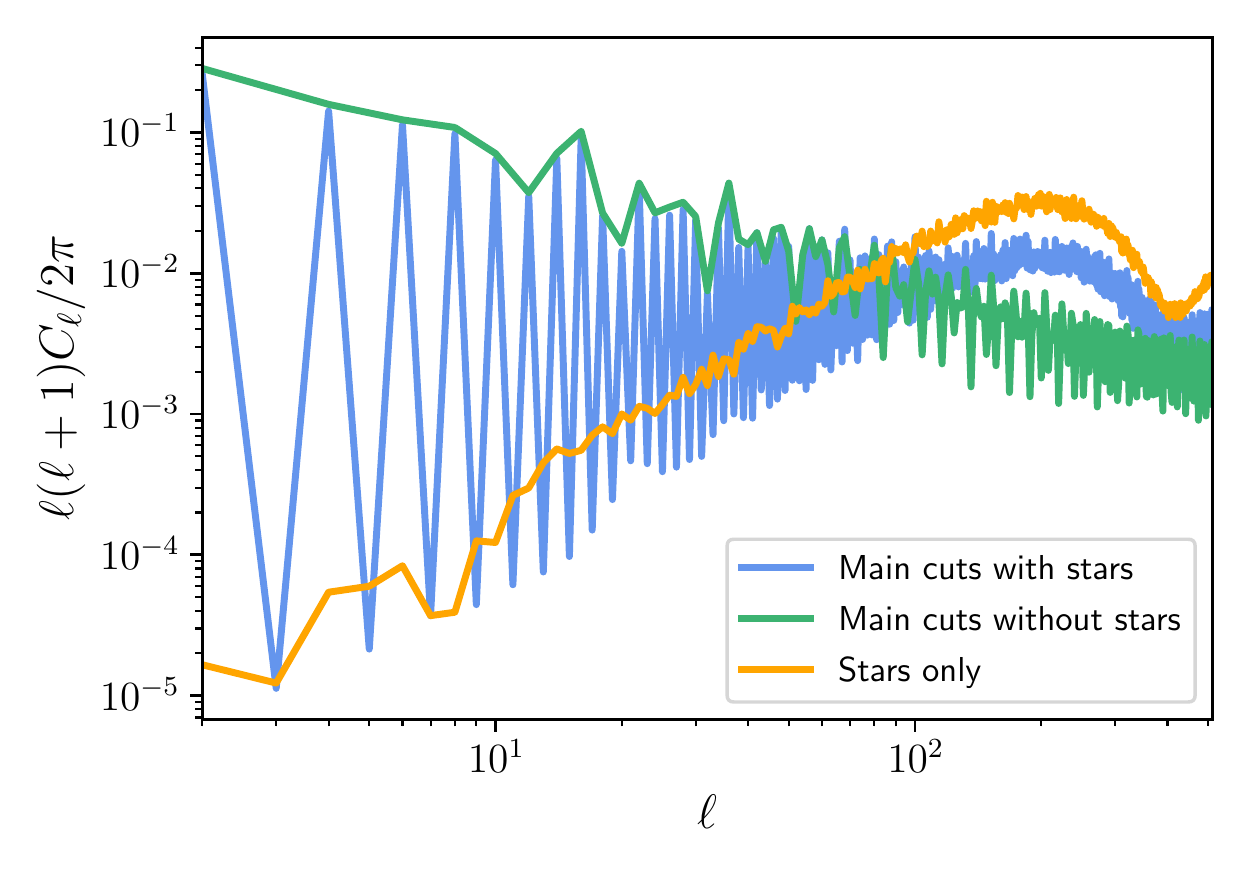}
    \vspace*{-0.75cm}
    \caption{Power spectrum of the sky masks used in our analysis. Note that for
      the `main cuts without stars' curve we plot the even $\ell$-modes only due
      to the very small values that odd $\ell$-modes take, arising from the
      parity of the mask.}
    \label{fig:Mask_power_spectrum}
\end{figure}

Since the exact form of Pseudo-$\Cl$ mixing matrix is highly sensitive to the
power spectrum of the mask used, through Equation~\ref{eqn:pcl_mixing_matrix},
we computed the spherical harmonic transform of
our generated mask. This is shown in Figure~\ref{fig:Mask_power_spectrum}.

We plot the power spectrum for our main galactic- and ecliptic- cuts only, the
two cuts with added star mask, and star mask only. For the case without stars
added, we plot the $\Cl$ values for even $\ell$-modes only. This is due to the
very small values for odd-$\ell$ modes arising from the parity of the mask.
Here, we see that the power spectrum for our main mask with stars added has two
distinct regions: dominated by the two main cuts for $\ell \lesssim 10^{2}$, and
dominated by the star mask above this threshold.

We can understand the primary behaviour of the mask's power spectrum through
computing the analytic power spectrum for a simple mask that is comprised of a
single horizontal cut ranging from $\theta = A$ to $\theta = B$. Doing so, we
find that the $\Cl$ values are given by
\begin{multline}
    \Cl = \frac{\pi}{\left(2 \ell + 1\right)^{2}} 
    [P_{\ell + 1}\left(\cos A\right) - P_{\ell - 1}\left(\cos A\right) + \\
    - P_{\ell + 1}\left(\cos B\right) + P_{\ell - 1}\left(\cos B\right)]^2.
    \label{eqn:Mask_Cl_theory_vals}
\end{multline}
Enforcing that the mask is symmetric around $\theta = \pi / 2$, and using the
parity of the Legendre polynomials, we find that the analytic prediction for the
odd $\ell$-modes are zero. The addition of a second cut of equal width, for
example the ecliptic-cut, keeps the reflective symmetry. While we use a slightly
thinner ecliptic-cut, we still keep this approximate symmetry. Propagating these
suppressed odd-$\ell$ modes into the mixing matrix through
Equation~\ref{eqn:pcl_mixing_matrix} explains the result for why we see strong
coupling in the covariance matrices between $\Cl$ values that have even-$\ell$
offsets, and little coupling between odd differences.

We also see that the amplitude of the power spectrum coefficients for the mask
without stars generally decreases at larger multipole values, where the $\Cl$
values roughly scale as $\ell ^2 \Cl \propto 1 / \ell$. This arises from the
large-$\ell$ behaviour of the Legendre polynomials, where they scale as
$P_{\ell} \propto 1 / \sqrt{\ell}$ \citep{Szego1975orthogonal}.

The behaviour of the power spectrum of the star mask can also be broadly split
into two distinct regions. The first is for multipoles $\ell \lesssim 200$ where
the $\Cl$ values are constant, which is a result of the random scatter of the
stars on the sphere resulting in a noise-like signal. The second is for
multipoles larger than $\ell \gtrsim 200$ where the $\Cl$ values start to
oscillate in a sinc-like behaviour, which is where the features of the
individual circular disks dominate.

\subsection[PseudoCl implementation]{Pseudo-$\Cl$ implementation}
\label{sec:PseudoCl_implementation}

An essential part of our work is the accurate computation of the $\Cl$ 
covariance matrices both for our new QML implementation and its comparison to
results obtained using the Pseudo-$\Cl$ method. Here, we used the 
\texttt{NaMaster}\footnote{\url{https://github.com/LSSTDESC/NaMaster}} code to
produce all estimates for the Pseudo-$\Cl$ method \citep{Alonso:2018jzx}.

In general, the computation of the exact Gaussian covariance is a difficult
problem that has been discussed extensively in previous literature. In our work,
we employed the narrow kernel approximation as presented in
\cite{Garcia-Garcia:2019bku} to compute this Gaussian approximation. However, it
should be noted that the narrow kernel approximation overestimates the variances
for the lowest $\ell$ multipoles. Since it is these exact multipoles that we are
most interested in, we instead opt to estimate the covariance from an ensemble
of $5\,000$ maps when investigating the raw variances. However, as the
estimation of the off-diagonal elements of the covariance matrix are highly
sensitive to the number of realisations in the ensemble, when we investigate
parameter constraints that are derived from the $\Cl$ covariance matrix, we use
the `analytic' result as returned from the narrow kernel approximation. In the
limit of using large numbers of realisations in the ensemble, at the cost of
extensive run-time, \cite{Garcia-Garcia:2019bku} demonstrated that these two
estimation techniques are consistent.

\section{Results}
\label{sec:Results}

\subsection{Benchmark against existing estimators}
\label{sec:QML_estimtors_benchmark}

\begin{figure}
  \includegraphics[width=\columnwidth]{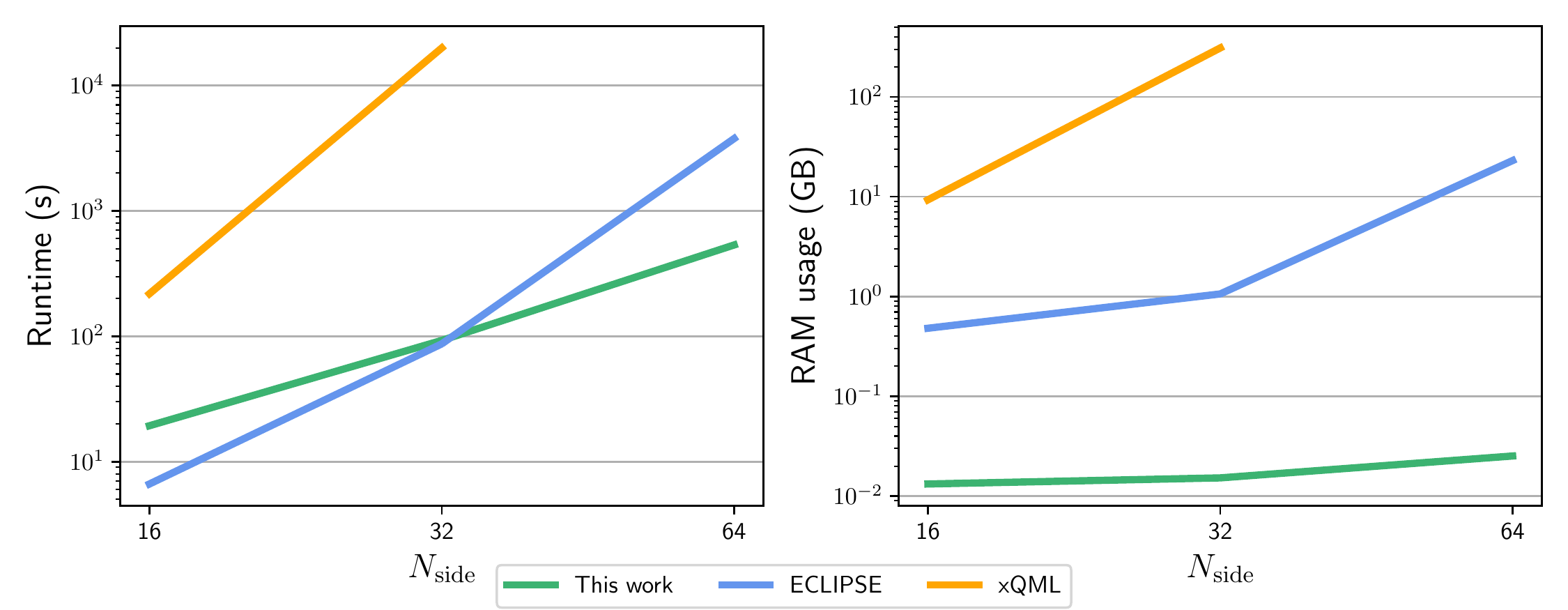}
  \vspace*{-0.5cm}
  \caption{Comparison of RAM usage and run-time for different implementations
  of the QML estimator. We show that our new method has significantly reduced
  RAM usage compared to existing estimators, which is why we can extend our
  method to increased map resolutions that the other methods cannot process.
  Results were obtained using an average over ten maps for our method, and 
  averaged over three runs for each method at each resolution. Computations
  were performed using 32 cores of an Intel Cascade Lake processor.}
  \label{fig:QML_code_comparison}
\end{figure}

In this section, we present the results of a comparative study between our new
QML implementation and the Pseudo-$\Cl$ method. First, we wish to investigate
how using the novel techniques employed by our new estimator impacts the ability
to recover the power spectrum when compared to existing QML implementations. We
compare with two leading public implementations of the QML estimator:
\begin{itemize}
    \item \texttt{xQML}\footnote{\url{https://gitlab.in2p3.fr/xQML/xQML}} 
    as presented in \cite{Vanneste:2018azc}. This is a
    straightforward implementation of the QML method as presented in 
    \cite{Tegmark:1996qt} and \cite{Tegmark:2001zv} that has been generalised
    to cross-correlations between maps. It is written primarily in Python with
    small parts written in C.
    \item \texttt{ECLIPSE}\footnote{\url{https://github.com/CosmoTool/ECLIPSE/}}
    as presented in \cite{Bilbao-Ahedo:2021jhn}. This is a more numerically
    efficient implementation of the QML estimator compared to the original
    prescription and thus exhibits somewhat better performance scaling with
    resolution over the naive method. It is written in \textsc{Fortran}.
\end{itemize}

We wish to compare the performance of our new code, written in \Cpp, with these
existing methods. In Figure~\ref{fig:QML_code_comparison}, we present a
comparison for the total run-time and RAM usage for the two codes described
above and our new method described in this work for a range of map resolutions
pushing to the highest $\Nside$ possible with these codes and the computational
resources available to us. Here, we see that while our new code is competitive
in total run-time when compared to \texttt{ECLIPSE}, we see many orders of
magnitude improvement in the total RAM usage for our method over the other two
methods. This is because we never have to explicitly store, invert, and compute
the product of any of the massive $\Npix \times \Npix$ matrices that the other
two methods employ. Since we are only interested in computing the action of the
covariance matrix on a trial pixel-space vector, we keep all of our
working-quantities as $\mathcal{O}(\Npix)$ which clearly have much better RAM
scaling with resolution over the pixel-space matrices. It is this vastly reduced
RAM usage requirement that allows us to push our method to resolutions that are
simply not possible on standard high performance clusters using the two
previously discussed methods. It is important to note that our new
implementation is now only run-time limited, and thus can be pushed to even
higher resolutions than have been considered in this work if more extensive
computing resources are available. The time-limiting steps to our implementation
is the transformations of the trial vector between pixel- and harmonic-space
through the use of the \texttt{HealPix} functions \texttt{alm2map} \&
\texttt{map2alm}. Since both of these functions are implemented using
\texttt{OpenMP} parallelism, faster run-times can be achieved through simply
running our code on higher core count processors.

Our code is publicly available and can be downloaded from
\href{https://github.com/AlexMaraio/WeakLensingQML}{\texttt{https://github.com/AlexMaraio/WeakLensingQML}
\faicon{github}}.

\subsection{Accuracy of numerical Fisher matrix}
\label{sec:Accuracy_of_numerical_fisher}

\begin{figure}
  \includegraphics[width=\columnwidth]{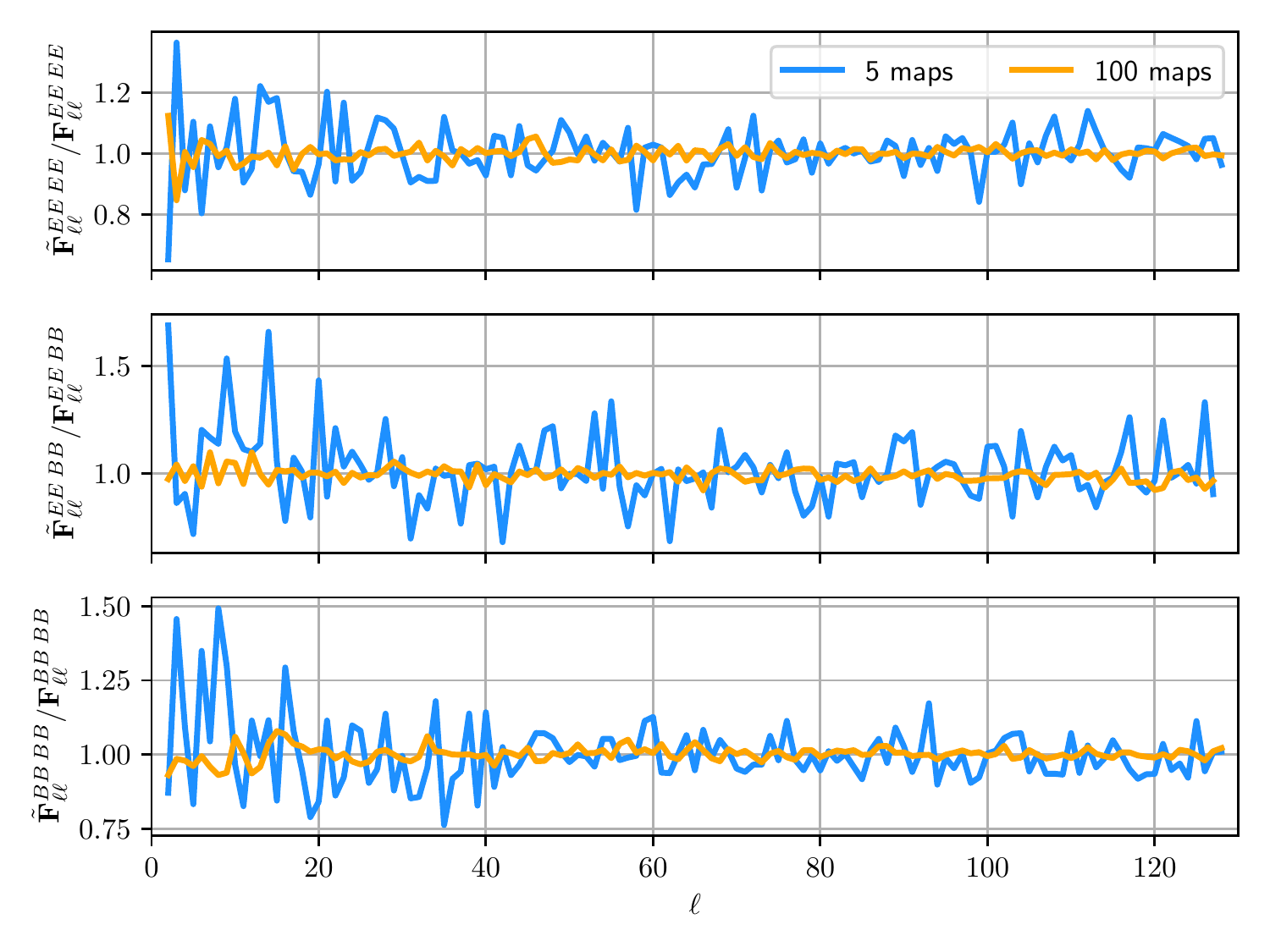}
  \vspace*{-0.75cm}
  \caption{Ratio of our numerical Fisher matrix with respect to the analytic
  result for a map resolution of $\Nside = 64$. Here, we plot the diagonal of
  the $EE$-$EE$, $EE$-$BB$, and $BB$-$BB$ components. All three components
  show good consistency with unity regardless of the number of maps averaged
  over when computing the Fisher matrix through
  Equation~\ref{eqn:FisherMatrixEstimation}, with the amplitude of the scatter
  decreasing with increasing number of maps.}
  \label{fig:Fisher_ratio}
\end{figure}

As explained in Section~\ref{sec:Finite_diff_fisher}, we use finite-differences
to compute the approximate form of the Fisher matrix from a set of $s_{\ell}$
values estimated using conjugate-gradient techniques. To validate the accuracy
of these methods we compared our estimates of the Fisher matrix with the
`analytic' result as computed using the formalism presented
in~\cite{Bilbao-Ahedo:2021jhn}. While this is still a `brute-force' QML
implementation, where the covariance matrix still needs to be computed and
stored in full, their method allows many quantities to be expressed in terms of
the spherical harmonic transform matrix $\mathbfss{Y}$ and thus reduce the
computational demands of the estimator. This comparison was performed at a map
resolution of $\Nside = 64$, which was the maximum resolution possible for the
computation of the analytic result. Note that at this resolution, the typical
pixel scale is larger than the angular size of our cut-outs generated for our
star mask, and so this comparison was computed for the case without the star
mask added to the main cuts. The result of this comparison is presented in
Figure~\ref{fig:Fisher_ratio}, where we plot the ratio of the diagonal of the
$EE$-$EE$, $EE$-$BB$, and $BB$-$BB$ components of the Fisher matrix. We plot the
cases for where we average over five and one hundred maps when injecting power
into the generated maps when estimating the Fisher matrix (see
Equation~\ref{eqn:FisherMatrixEstimation}).
Figure~\ref{fig:Numeric_to_analytic_Fisher_ratio_grid} shows this ratio extended
to several of the off-diagonal strips, specifically for the cases with $\Delta
\ell = 2, 8, 32$. Both figures show that while the amplitude of the random
scatter in the ratio decreases significantly when averaging over more maps, both
cases are simply random scatter around unity - and thus our numerically obtained
Fisher matrix is a true representation of the actual Fisher matrix. Propagating
the two different numerical and analytical $\Cl$-Fisher matrices to parameter
constraints using Fisher forecasts shows negligible differences in parameter
contours which again highlights our trust in our new method to estimate the
$\Cl$-Fisher matrix at any resolution. Hence, we are free to use our validated
method to reliably increase the resolution of our implementation beyond what is
possible with current implementations. In our analysis, we averaged over twenty
five random realisations which provided a good compromise between run-time and
numerical accuracy.

\subsection[Comparing Cl variances of QML to PseudoCl]{Comparing $\bm \Cl$ variances of QML to Pseudo-$\Cl$}

With our new implementation, we can extend the analysis of the properties of the
QML estimator to map resolutions of $\Nside = 256$, which allows us to
accurately recover the power spectrum up to a maximum multipole of $\lmax = 512$.
At this resolution, the storage of the full pixel covariance matrix alone would
require approximately $5\,\mathrm{TB}$ of RAM, which is clearly an unfeasible
requirement for any current computer and thus any analysis at this resolution is
not achievable using current QML implementations. 

We generate maps of the weak lensing shear through producing Gaussian
realisations with the power spectrum described in
Section~\ref{sec:Theory_power_spectrum} (see
Figure~\ref{fig:FiducialPowerSpectrum}). We then add shape noise to these maps
according to Equation~\ref{eqn:Noise_power_spectrum}. We generate twenty five
such maps for each power spectrum multipole that we are injecting power into
($EE$ and $BB$, from $\ell = 2$ to $\ell = 767$) and compute the QML power
spectrum ($EE$ and $BB$) from each. We estimate the QML covariance matrix using
the fact that for Gaussian maps the inverse Fisher matrix is the covariance
matrix. The methods described in Section~\ref{sec:Finite_diff_fisher} were used
to estimate the Fisher matrix, where we average over twenty five maps per
multipole when evaluating Equation~\ref{eqn:FisherMatrixEstimation}.
The Pseudo-$\Cl$ covariance matrix was computed using the methods described in
Section~\ref{sec:PseudoCl_implementation}, and was constructed from an ensemble
of $5\,000$ maps. We do not bin in $\ell$ either of our QML or Pseudo-$\Cl$
estimators, noting that our mask is small enough that the Pseudo-$\Cl$ mixing
matrix is invertible without binning as we are able to reconstruct all modes
that we have generated. We note that our results were robust to different
binning strategies that were applied though not used in our final results. 

\begin{figure}
    \includegraphics[width=\columnwidth]{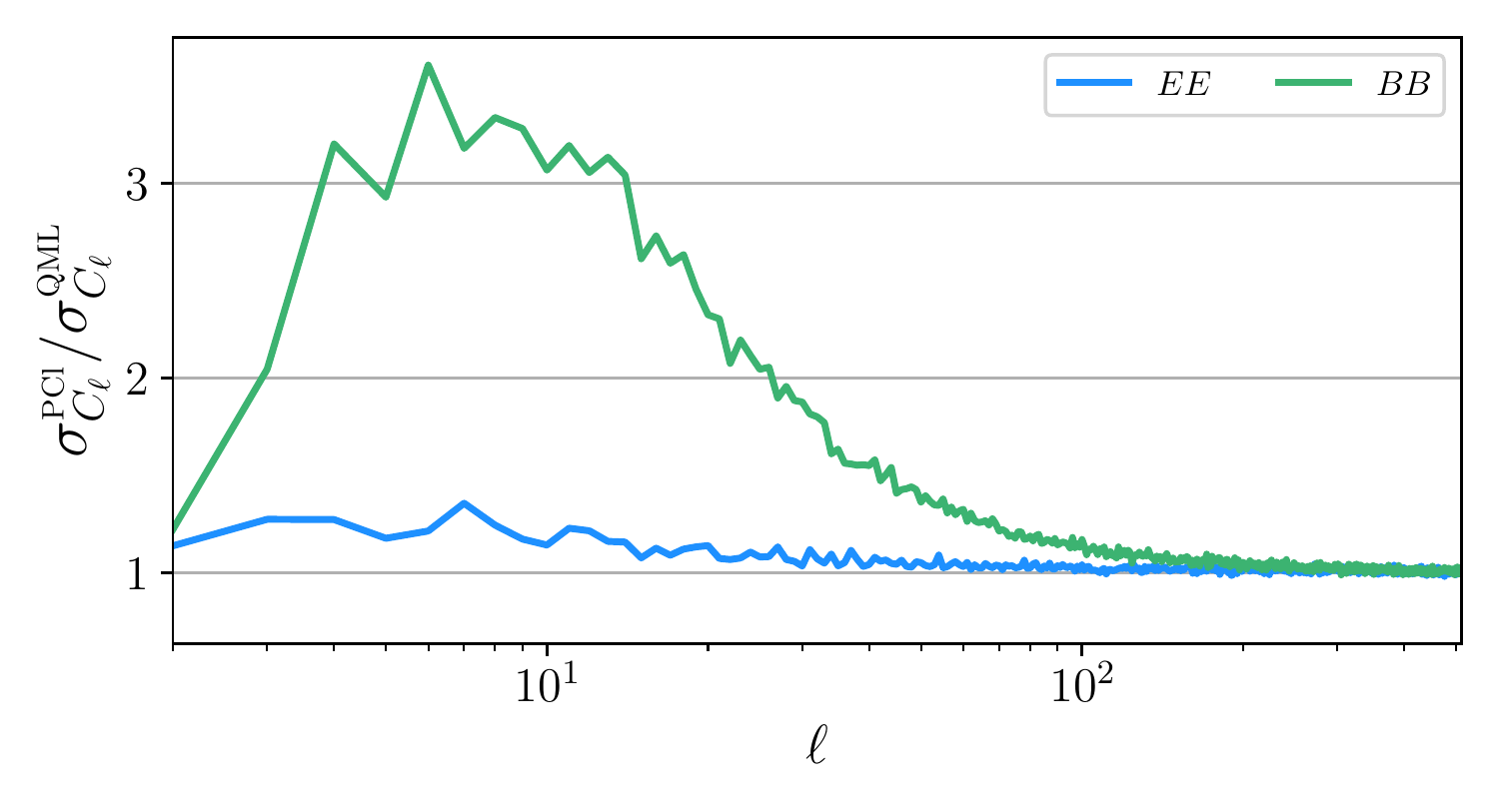}
    \vspace*{-0.75cm}
    \caption{Ratio of standard deviation of the $\Cl$ values (top curve $\ClBB$,
      bottom curve $\ClEE$) obtained using the deconvolved Pseudo-$\Cl$ method
      using \texttt{NaMaster} to those obtained using our new QML
      implementation. We see that the QML estimator provides the largest
      improvements over the Pseudo-$\Cl$ method on the largest angular scales,
      with a very significant improvement for the $B$-modes.}
    \label{fig:Cl_err_ratio}
\end{figure}

In Figure~\ref{fig:Cl_err_ratio} we plot the ratio of the standard deviations
associated with the Pseudo-$\Cl$ estimator with respect to our QML
implementation for the diagonal values associated with the $EE$-$EE$ and
$BB$-$BB$ block of the covariance matrix. Here, we see that the Pseudo-$\Cl$
estimator is sub-optimal to the level of $\sim\!\!20\%$ for $\ell \lesssim 50$
for the $EE$ spectra. This corresponds to an equivalent increase in the survey
area of around $40\%$ on these scales, which is a massive increase in equivalent
area considering that forthcoming Stage-IV surveys are expected to maximise the
possible sky area that is observable for ground- or space-based cosmic shear
surveys~\citep{Euclid:2021icp}. Hence, getting this additional area `for free'
by analysing the data through QML methods demonstrates the advantages of using
such methods and such investigations into their behaviour for cosmic shear
analyses. For the $BB$ spectra, we find that the Pseudo-$\Cl$ estimator produces
errors that are many times that of the optimal QML estimator, peaking at over
three times the standard deviation for the Pseudo-$\Cl$ estimator with respect
to our QML method. This ratio remains significantly above unity for multipoles
that are well above one hundred, which shows that there is a huge advantage to
be gained in $B$-mode precision when using QML methods over the Pseudo-$\Cl$
estimator. We find that the ratio for both sets of spectra decays to unity (with
some random scatter) for larger $\ell$ values. This matches previous QML
studies, which have principally been conducted in the context of the CMB and
ground-based galaxy clustering surveys, which found that the Pseudo-$\Cl$
estimator is close to optimal on small scales and for homogenous noise, and we
find similar results here in the weak lensing
context~\citep{Efstathiou:2003dj,Leistedt:2013gfa}. 

Figure~\ref{fig:Cl_err_ratio} also shows that the statistical precision of the
$B$-mode power spectrum is significantly higher for the QML method compared with
Pseudo-$\Cl$ method. This is relevant because cosmic shear theory predicts zero
$\ClBB$ modes and so any detection of a non-zero $\ClBB$ signal would prompt a
thorough investigation of the data \citep{Kilbinger:2014cea}. Potential sources
of $B$-mode power include residual point-spread function uncertainties,
telescope detector defects, and intrinsic alignments, all of which should be
investigated if non-zero $B$-mode power was found to be statistically
significant. The Pseudo-$\Cl$ estimator is very sub-optimal on large scales, and
so this loss of sensitivity to the $B$-modes can arise from contamination
leakage from the $E$-modes into the $B$-modes due to the nature of the cut-sky.
Since the QML estimator is derived from the likelihood for the maps, which
depends on the input fiducial power spectrum which contains zero $B$-mode power,
any $B$-mode power present in the masked maps must arises from leakage from the
$E$-modes, and thus the estimator can weight the data optimality through the
covariance matrix to minimise the variance from the $E$-modes contributing to
the $B$-modes. For the Pseudo-$\Cl$ estimator, this leakage can be mitigated
through the map-level procedure of $B$-mode purification
\citep{Lewis:2001hp,Smith:2005gi,Grain:2009wq} and has been shown to decrease
dramatically the associated $B$-mode errors, particularly at low $\ell$
multipoles~\citep{Alonso:2018jzx}. However, a requirement for purification to
work is that the mask must be differentiable along its edges. This can be
achieved through the apodisation of the mask which convolves the mask with some
smoothing window function that ensures differentiability. This has most commonly
been applied to cosmic microwave background experiments where their masks are
generally formed of a single much simpler cut applied to the sky
\citep{Planck:2018yye}. This allows apodisation to work effectively on the mask
without significant reduction to $\fsky$. However as previously discussed, a
weak lensing experiment also needs to mask out small regions corresponding to
bright stars or other objects that need removing from the data. These small
regions cause significant issues with the apodisation process as the convolution
with the smoothing function serves to dramatically increase their apparent area
- producing a significant reduction in $\fsky$. We investigate the effects of
apodising our mask in Appendix~\ref{app:Apodising_mask}. We find that while
apodisation strongly reduces widely separated mode coupling arising from the
suppression in small-scale power of the mask (as now the shape edges from our
star-like disks are smoothed out), this could not offset the significant
reduction in sky area (from $\fsky = 33\,\%$ to $\fsky = 22\,\%$) that
apodisation brought about. This resulted in apodisation providing looser
parameter constraints than for the case without apodisation. A key advantage of
the QML estimator is the natural $E$/$B$-mode separation without a loss in
sky area \citep{Bunn:2016lxi}.



\subsubsection{Varying noise levels}
\label{sec:Varying_noise_levels}

\begin{figure}
    \centering
    \includegraphics[width=\columnwidth]{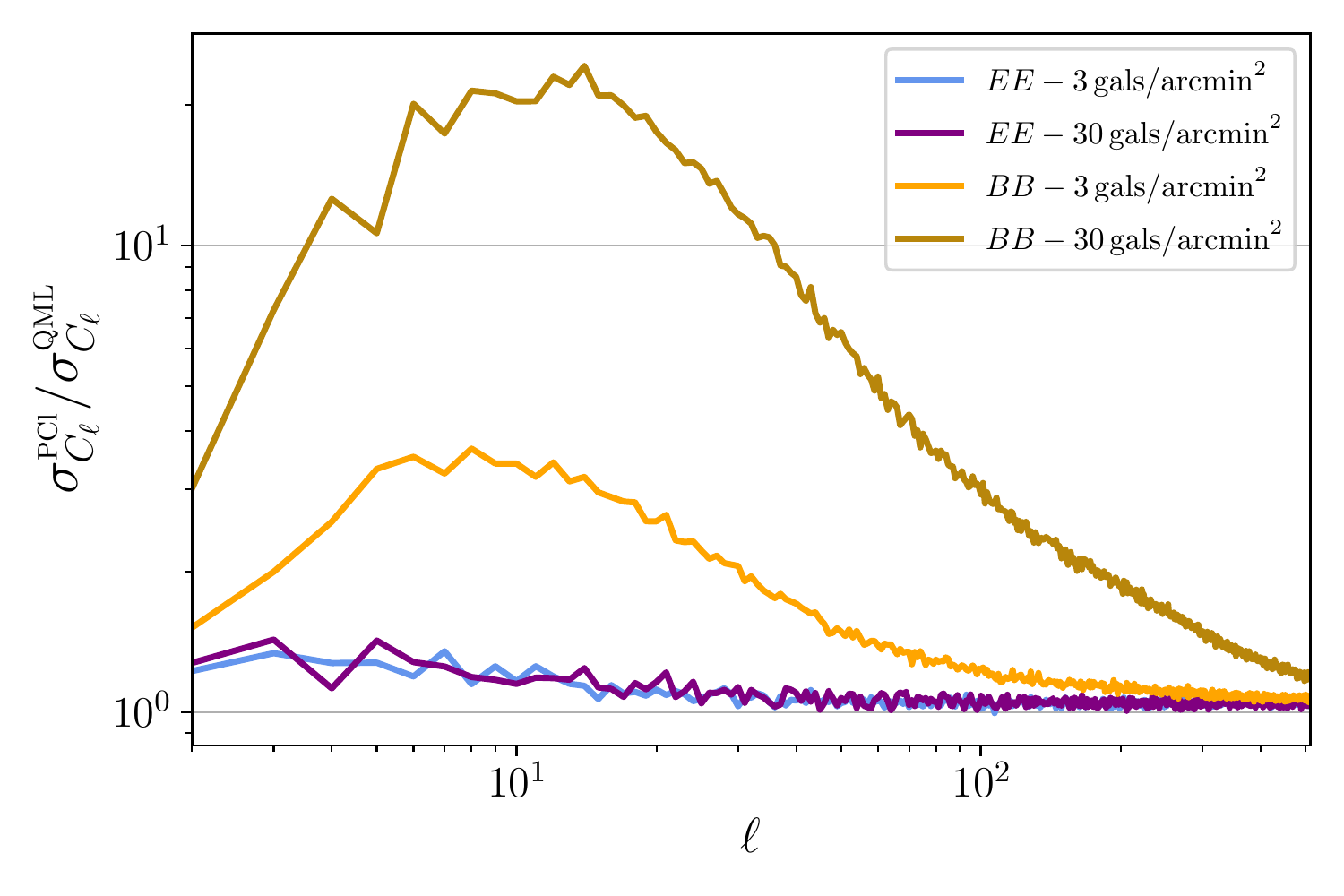}
    \vspace*{-0.75cm}
    \caption{Ratio of the errors on the power spectrum for the Pseudo-$\Cl$
      estimator with respect to the QML estimator for the case of two different
      noise levels of $\bar{n} = 3\,\textrm{gals / arcmin}^{2}$ and $\bar{n} =
      30\,\textrm{gals / arcmin}^{2}$. Here, we see the decreased noise level
      for the curves for the case of $\bar{n} = 3\,\textrm{gals / arcmin}^{2}$
      has negligible effect on the errors associated with the $EE$-spectra,
      whereas there is a large increase in the ratio for the $BB$-spectra.}
    \label{fig:PCl_to_QML_stdev_ratio_noise}
\end{figure}

\begin{figure}
    \includegraphics[width=\columnwidth, trim={0cm 0cm 0cm 0cm}, clip]{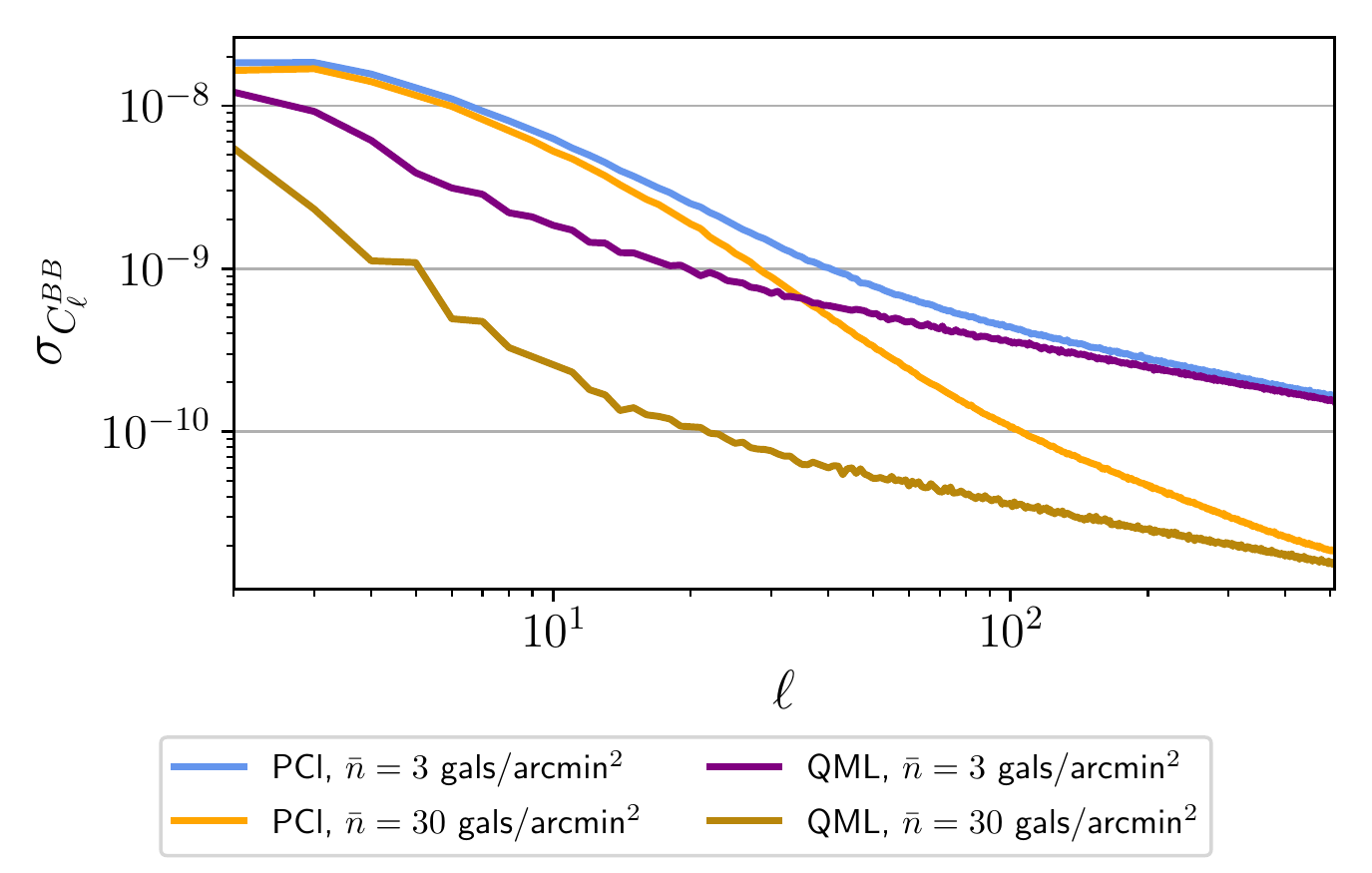}
    \vspace*{-0.6cm}
    \caption{Errors on the $BB$-spectra for the two different noise levels
      considered. Here, we see that the errors for the Pseudo-$\Cl$ estimator
      remain relatively unchanged for the lowest multipoles, whereas the errors
      for the QML estimator decrease dramatically when the amplitude of the
      noise is reduced. }
    \label{fig:Cl_std_BB_noise}
\end{figure}

Thus far, we have assumed a noise level corresponding to an experiment that has
observed thirty galaxies equally divided into ten redshift bins, giving $\bar{n}
= 3\,\mathrm{gals / arcmin}^{2}$. Here, we wish to investigate the statistics
of our estimators for the case where the observed galaxies are combined into a
single redshift bin, giving a much lower noise level of $\bar{n} =
30\,\mathrm{gals / arcmin}^{2}$. Performing this comparison produced results as
shown in Figure~\ref{fig:PCl_to_QML_stdev_ratio_noise}. Here, we see that while
there is negligible differences in the relative statistics between the
$E$-modes, there was a large increase in ratio between our two estimators for
the $B$-modes. We can investigate this large difference in the $B$-mode ratios
by plotting the raw errors of the $B$-mode spectra for our two estimators for
the two noise cases, which is shown in Figure~\ref{fig:Cl_std_BB_noise}.

Here, we see that for low multipoles there is a large difference in the errors
for the QML estimator between the two noise levels whereas the errors for the
Pseudo-$\Cl$ estimator remains relatively unchanged. At this decreased noise
level, the noise is subdominant to the cosmic variance from the $E$-modes on
large-scales. Hence, the QML estimator can vastly outperform the Pseudo-$\Cl$
method on these large-scales because the QML likelihood can efficiently minimise
the cosmic variance from the $E$-modes leaking into the $B$-modes through the
cut-sky. The increased noise level corresponds to a genuine increase in $B$-mode
power, which the QML estimator cannot suppress as efficiently as the $E$-mode
cosmic variance leakage, and so we see an associated decrease in the ratio
between its errors and that of the Pseudo-$\Cl$ estimator.

\subsection{Cosmological parameter inference}
\label{sec:Parameter_contours}

\begin{figure}
    \includegraphics[width=\columnwidth, trim={0cm 0cm 0cm 0cm}, clip]{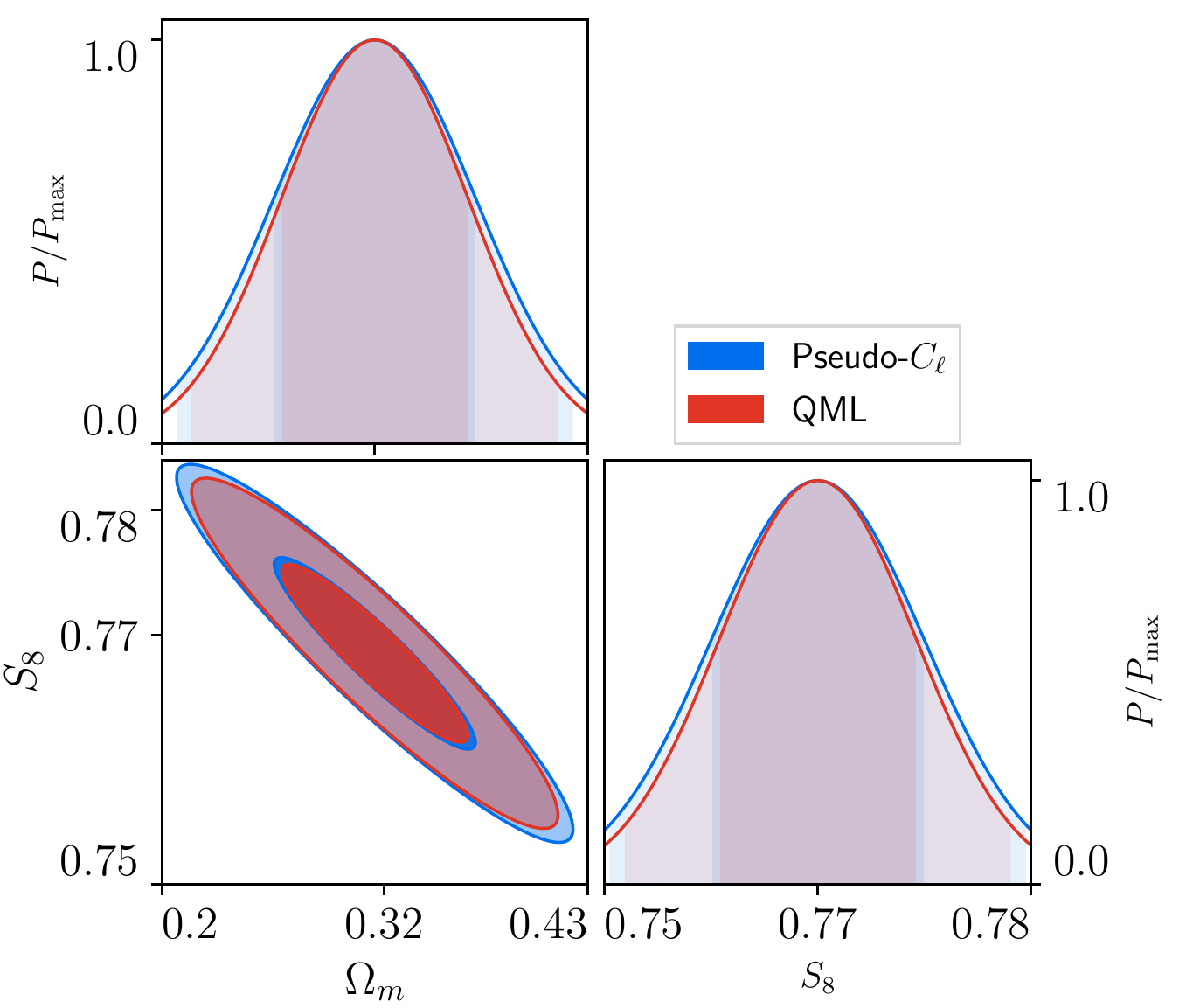}
    \vspace*{-0.5cm}
    \caption{Parameter constrains on $\Seight$ and $\Omegam$ obtained from a Fisher
      matrix analysis up to a maximum multipole of $\lmax = 512$ for our two
      estimators. Here, we see that the increased errors associated with the
      Pseudo-$\Cl$ method propagate through to slightly broadened parameter
      contours.}
    \label{fig:Param_comparison}
\end{figure}

A Fisher matrix forecast was used to propagate our estimated $\Cl$ covariance
matrices into parameter constraints. For an arbitrary set of cosmological
parameters $\vartheta_{\alpha}$ and $\vartheta_{\beta}$, the corresponding
Fisher matrix element is given by \citep{Tegmark:1996bz}
\begin{equation}
    \mathcal{F}_{\alpha \beta} = 
    \sum_{\ell, \, \ell'}
    \frac{\partial C_{\ell}}{\partial \vartheta_{\alpha}} \, 
    \mathbfss{C}^{-1}_{\ell \ell'} \, 
    \frac{\partial C_{\ell'}}{\partial \vartheta_{\beta}},
\end{equation}
where $\mathbfss{C}$ is the $\Cl$ covariance matrix. In our analysis, we focused
on the two parameters that cosmic shear places the tightest constraints on: the
clustering amplitude $\Seight$ and the total matter density $\Omegam$.
Figure~\ref{fig:Param_comparison} shows a comparison of the derived constraints
for these two parameters between our two estimators. Here, we see the effect of
the slightly increased errors associated with the Pseudo-$\Cl$ estimator have
propagated into slightly increased contours for these two parameters when
compared to the QML estimator's contours. This result could have been
anticipated from Figure~\ref{fig:Cl_err_ratio}, since most of the information on
these parameters originates from small scales where the ratio of the errors
approaches unity. In contrast, parameters affecting large angular scales, such
as local primordial non-Gaussianity, are expected to benefit substantially from
an optimal method.

\begin{figure}
    \includegraphics[width=\columnwidth]{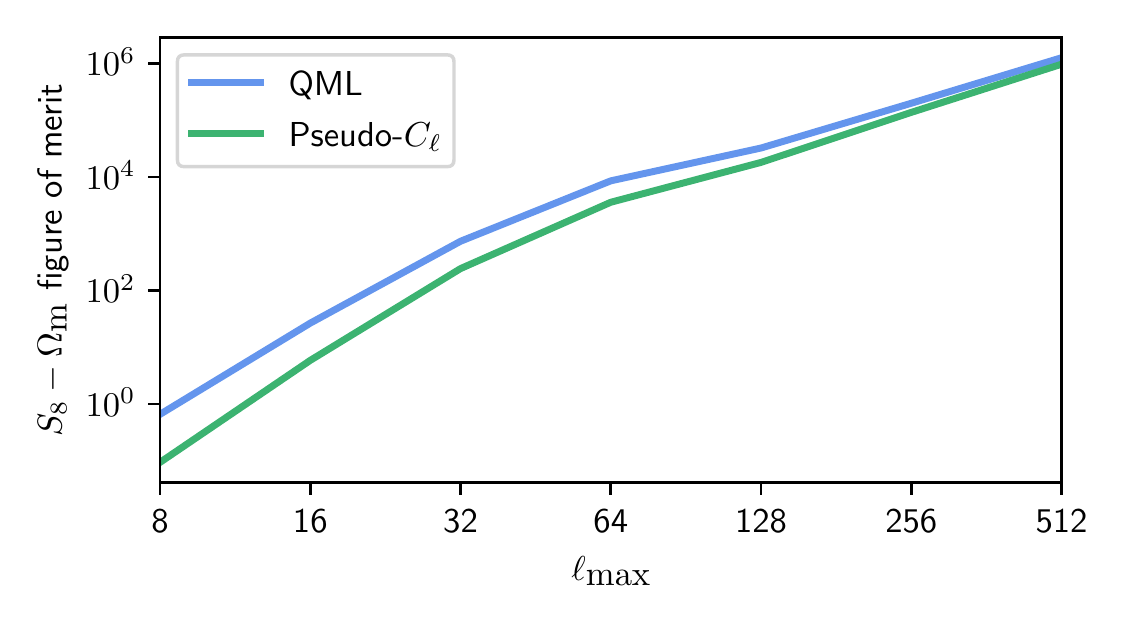}
    \vspace*{-0.75cm}
    \caption{Values for the figure of merit for the combination of $\Seight$ and
      $\Omegam$ as a function of the maximum $\ell$ multipole used in the
      analysis. We see that as the maximum multipole increases, the relative
      sub-optimality of the Pseudo-$\Cl$ estimator decreases and results from
      the two estimators converge.}
    \label{fig:figure_of_merit_vs_lmax}
\end{figure}

The figure of merit, which quantifies how well constrained parameters are, is
related to the Fisher matrix through~\citep{Euclid:2019clj} 
\begin{equation}
    \mathrm{FoM}_{\Seight \Omegam} = \sqrt{\mathrm{det}\left(\mathcal{F}\right)}.
\end{equation}
A plot of the figure of merit for the combination of $\Omegam$ and $\Seight$ as
a function of maximum multipole is shown in
Figure~\ref{fig:figure_of_merit_vs_lmax}. Here, we see that the sub-optimality
of the Pseudo-$\Cl$ method is most apparent when we are limited to low
multipoles. As the maximum multipole increases we see that the figures of merits
converge, however showing that the QML method consistently out performs the
Pseudo-$\Cl$ method. 

The application of Fisher forecasting to predict parameter constraints from the
covariance matrix is done under the assumption that the $\Cl$ values recovered
from the estimators can be described by a Gaussian likelihood. While it has been
shown that for the full sky case, an analytic calculation of the likelihood of
the power spectra can be computed \citep{Hall:2022das}, which can be accurately
modelled as a Gaussian on small scales, the exact likelihood of the recovered
power spectrum using either the QML or Pseudo-$\Cl$ estimators is still unknown.
Previous works have simply used the Gaussian approximation citing the central
limit theorem \citep{Seljak:2017rmr}. Since QML estimators produce estimates for
the full sky power spectrum, we expect this to follow a Gaussian distribution
more closely than the cut-sky estimates produced from Pseudo-$\Cl$ methods and
so this approximation is expected to hold better for QML over Pseudo-$\Cl$
methods.

\subsubsection{Inclusion of stars}

Our results presented thus far have been all for the case where we have applied
a star mask to a large-scale mask featuring ecliptic and galactic cuts, as shown
in Figure~\ref{fig:SkyMaskWithStars}. Here, we wish to investigate the detailed
effects on the covariances of our estimators when we apply the star mask to our
main two cuts. All results here are presented for the case without any mask
apodisation applied.

\begin{figure}
    \includegraphics[width=\columnwidth]{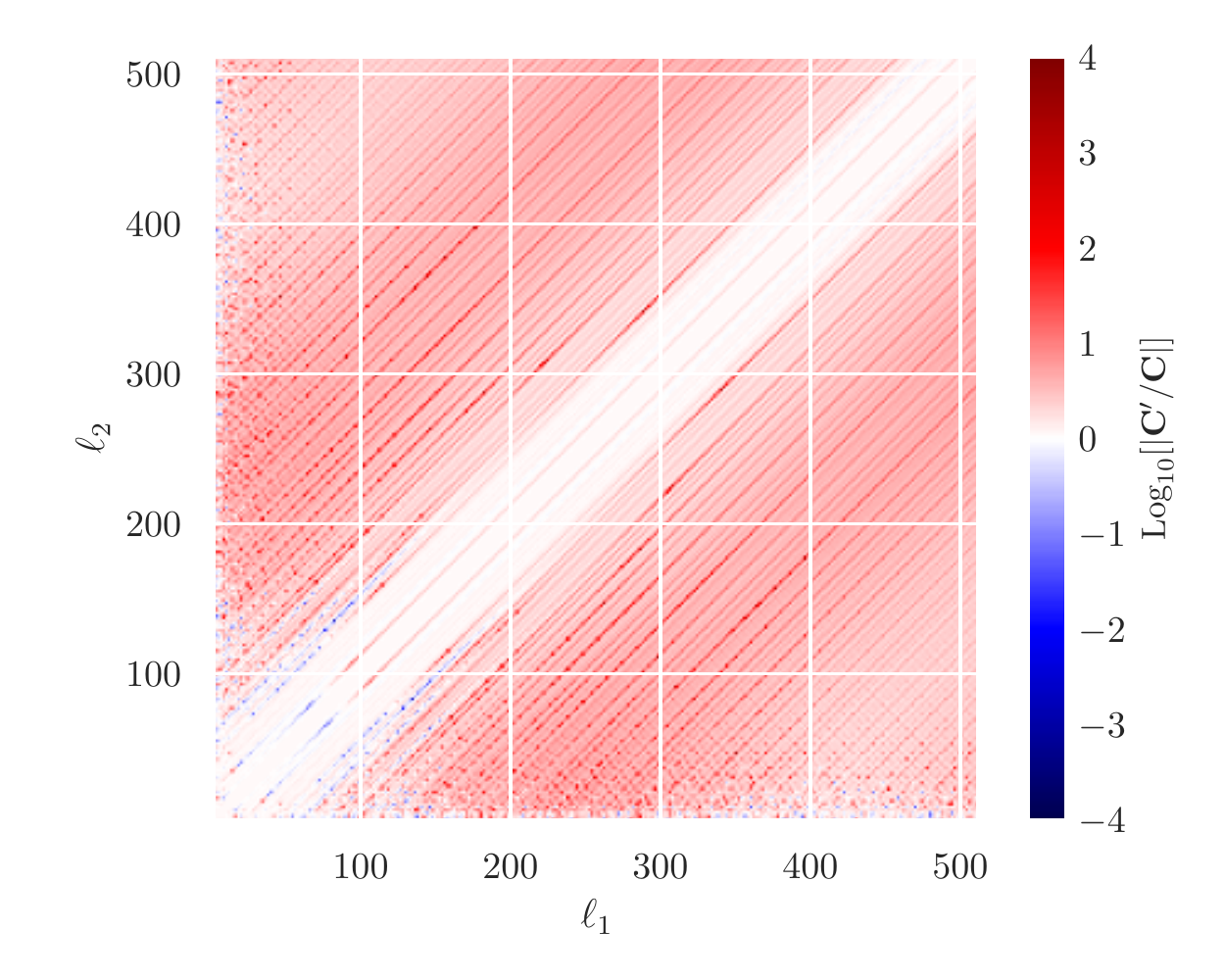}
    \vspace*{-0.5cm}
    \caption{Ratio of the analytic Pseudo-$\Cl$ covariance matrix for the
      $\ClEE$ power spectrum for the cases with ($\mathbfss{C}^{\prime}$) and
      without stars ($\mathbfss{C}$) applied to the main mask. Note that we only
      plot the covariance matrix for even-$\ell$ values only (due to the very
      small values for the odd-$\ell$ case and so their ratios are dominated
      by numerical noise).}
    \label{fig:PCl_cov_star_ratio}
\end{figure}

The ratio of the Pseudo-$\Cl$ covariance matrix for the cases with and without
stars is presented in Figure~\ref{fig:PCl_cov_star_ratio}. Here, we see that the
primary effect of the addition of stars into the mask is to increase the
correlation between widely separated $\ell$-modes, while leaving the values close to
the diagonal in the covariance matrix relatively unchanged.

\begin{figure}
    \includegraphics[width=\columnwidth, trim={0cm 0cm 0cm 0cm}, clip]{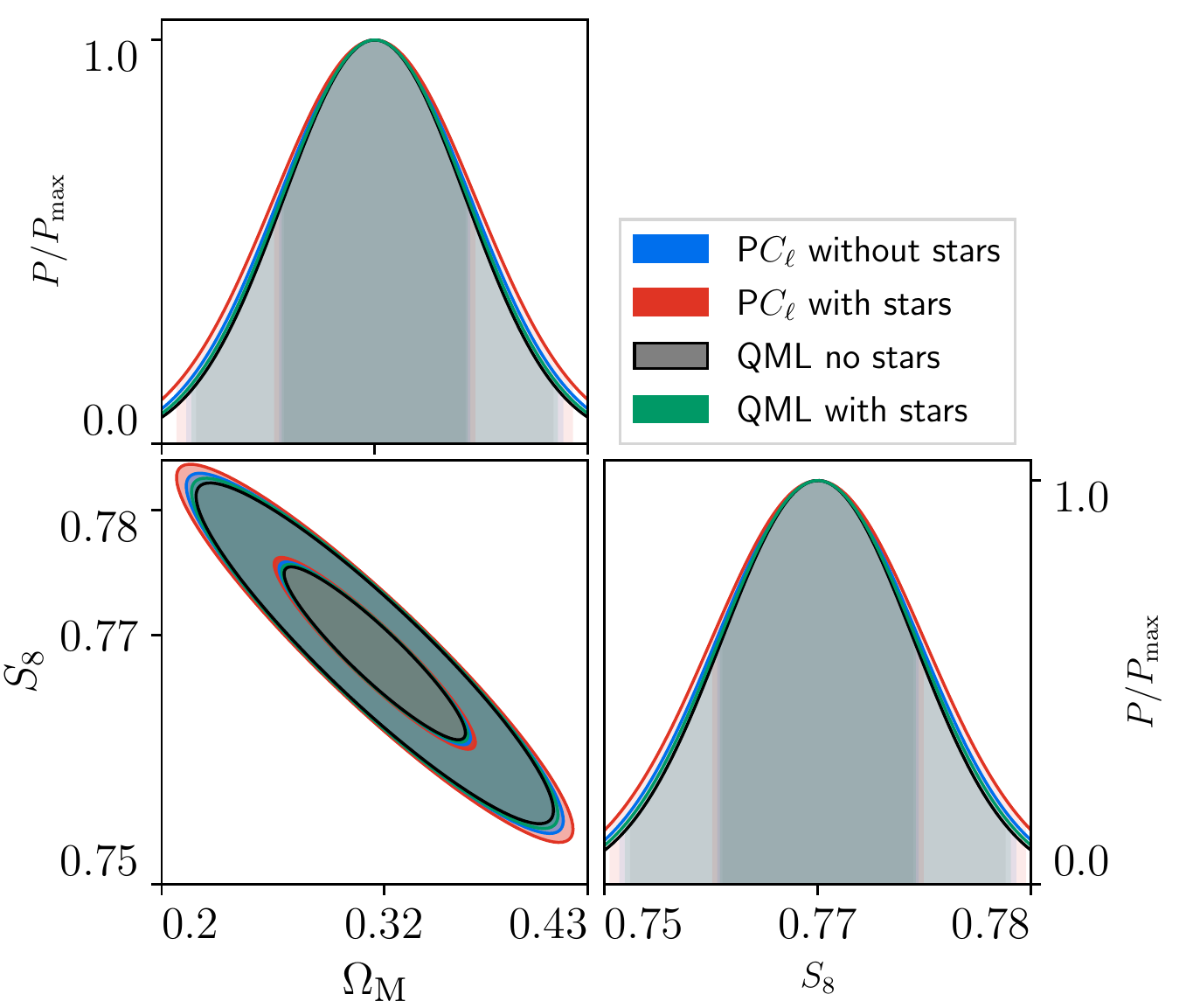}
    \vspace*{-0.5cm}
    \caption{Fisher parameter constraints comparison between QML and
      Pseudo-$\Cl$, where both estimators have a maximum multipole of $\lmax =
      512$, for the cases with and without the star mask applied to both
      estimators. Here we see that the Pseudo-$\Cl$ estimator is more sensitive
      to the inclusion of the star mask through the relative increase in
      parameter contours when compared to the QML contours.}
    \label{fig:Parameter_constraints_stars}
\end{figure}

This new covariance matrix can then be propagated into parameter contours to see
if these increased long-range correlations (which fiducially have very small
values) have any meaningful effect on cosmological parameter constraints. This
is shown in Figure~\ref{fig:Parameter_constraints_stars}. Here, we see that
there are negligible differences on the parameter contours between the two cases
for our QML estimator, however there is a slight broadening in the contours for
the Pseudo-$\Cl$ estimator which is consistent with the loss of sky area to the
star mask. This shows that the Pseudo-$\Cl$ estimator is more sensitive to the
addition of a star mask than the QML estimator, further highlighting the
benefits of the QML method. 

\subsection{Non-Gaussian maps}
\label{sec:Non_gaussian_maps}

Throughout our paper, we have been applying our estimator to Gaussian
realisations of the cosmic shear field. However, as it has been shown that the
convergence field $\kappa$ is more accurately described by a log-normal
distribution \citep{Taruya:2002vy,Hilbert:2011xq}, an investigation of how our
estimators perform when applied to these non-Gaussian maps was undertaken. This
is because as the QML estimator assumes that the underlying power spectrum
coefficients follow a Gaussian distribution, any non-Gaussianities present
in the shear field could induce non-optimality into the recovered power spectra
which would increase errors.

\begin{figure}
    \includegraphics[width=\columnwidth, trim={0cm 0cm 0cm 0cm}, clip]{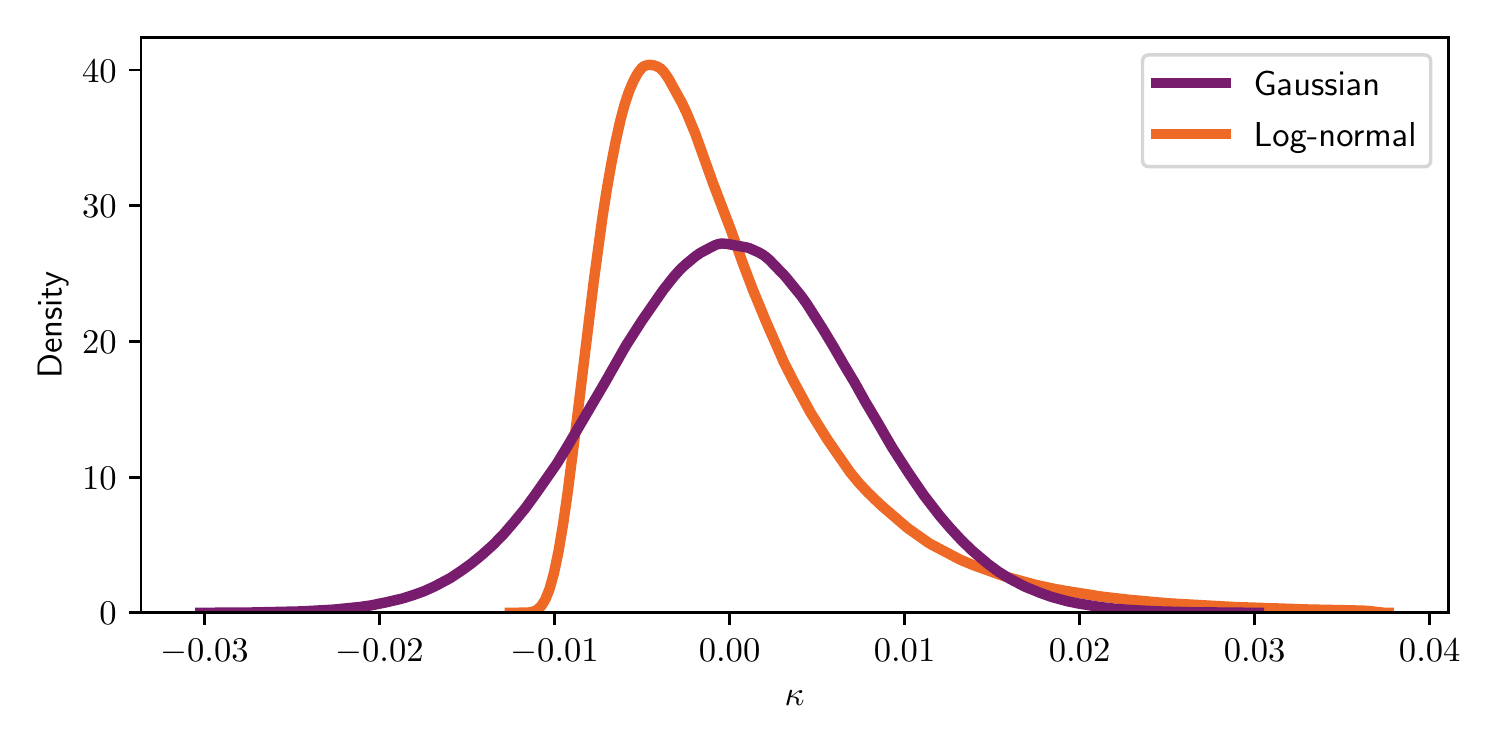}
    \vspace*{-0.5cm}
    \caption{Histogram showing the field values of a Gaussian and log-normal 
      realisation of the convergence field $\kappa$ for the same underlying
      power spectrum}
    \label{fig:kappa_histogram}
\end{figure}

The \texttt{Flask} software package was used to generate log-normal
maps~\citep{Xavier:2016elr}. The `shift parameter' that corresponds to the
minimum value of the convergence field required by \texttt{Flask} was set to
0.01214, following \cite{Hall:2022das}. The log-normal maps were generated at a
resolution of $\Nside = 1024$ and then downgraded to a resolution of $\Nside =
256$ as required to be processed through our estimators. A histogram showing the
distribution of $\kappa$ values for a Gaussian and log-normal realisation is
shown in Figure~\ref{fig:kappa_histogram}. This log-normal convergence field is
then propagated to a slightly modified shear field, which we can apply the
estimators to.

\begin{figure}
    \includegraphics[width=\columnwidth, trim={0cm 0cm 0cm 0cm}, clip]{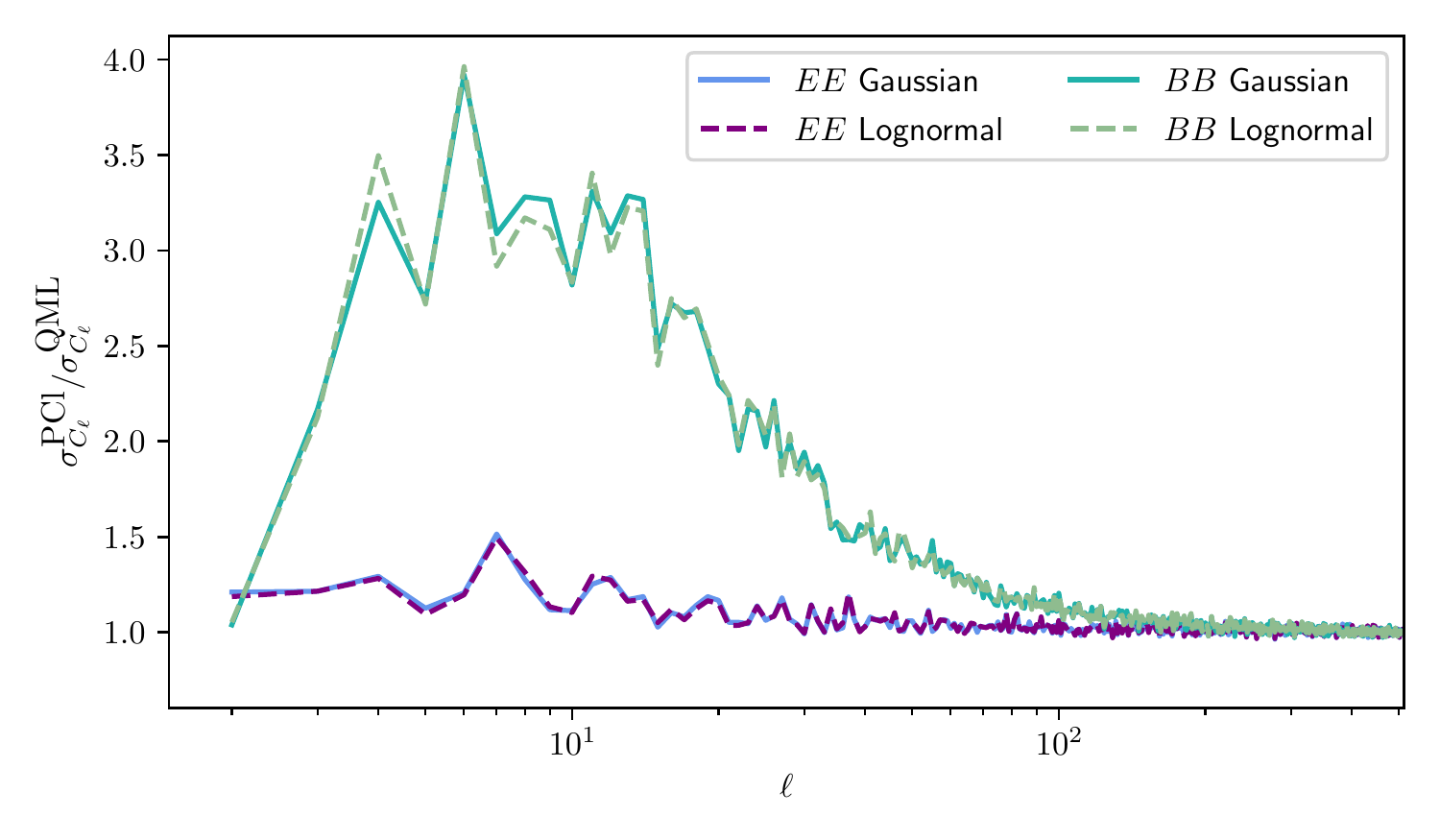}
    \vspace*{-0.5cm}
    \caption{Ratio of the power spectrum errors for the Pseudo-$\Cl$ method with
      respect to the QML estimator applied to both Gaussian and log-normal
      realisations. Here, we see that the results are indistinguishable between
      the two distributions. }
    \label{fig:Cl_std_ratio_nongaussian}
\end{figure}

Since the covariance of the QML estimator is no longer given by the inverse
Fisher matrix, we have to obtain estimates for the QML errors from an ensemble
of numerical realisations. The results of applying the QML and Pseudo-$\Cl$
estimators to an ensemble of $2\,500$ Gaussian and log-normal realisations is
shown in Figure~\ref{fig:Cl_std_ratio_nongaussian}. Here, we see that the ratio
of the power spectrum errors for the two distributions are virtually identical
which demonstrates that the relative behaviour of our estimators remains
unchanged even when applied to maps that the underlying likelihood does not
fully describe. Since the largest differences between the estimators for our
setup occur on the largest scales which is where the effects of the
non-Gaussianity is weak, this is not surprising.

\section{Conclusions}
\label{sec:Conclusions}

We have presented a new implementation of the optimal quadratic maximum
likelihood estimator that is the most efficient publicly available code of its
type. Using our new estimator, we have compared the statistical properties of
the expected power spectrum for forthcoming Stage-IV weak lensing surveys, using
realistic survey conditions, between our QML implementation and an existing
Pseudo-$\Cl$ code. We found that the sub-optimality of the Pseudo-$\Cl$
estimator resulted in marginally increased statistical errors for the $E$-mode
power spectra propagating to increased parameter contours when using Fisher
forecasting. In addition, we found a significant increase in the precision for
the $B$-mode power spectra when applying our QML estimator over the Pseudo-$\Cl$
method, which raises the hopes of being able to further constrain new $B$-mode
physics using forthcoming surveys. Our results show that the application of QML
methods to cosmic shear data provides a useful cross-check to existing methods,
and could have many interesting applications for the constraints of new $B$-mode
physics.

Our new estimator could be extended in numerous ways, for example the QML method
can be easily applied to spin-0 fields such as photometric galaxy clustering.
Since our estimator yield the best improvements on the largest physical scales,
scales at which primordial non-Gaussianity has the largest effects on the
observed signal, the use of our new estimator could enable tighter constraints
on primordial non-Gaussianity. In addition, photometric galaxy clustering data
can be combined with cosmic shear to form a combined $3 \times 2$-point
investigation which our estimator could be applied to. We leave these
applications of our estimator to future work. We also note that our estimator
could also be applied to thermal Sunyaev-Zeldovich data, which is a powerful
probe of cosmology at relatively low multipoles ($\ell \lesssim
10^{3}$)~\citep{Horowitz:2016dwk,Bolliet:2017lha}. This overlaps with the
multipole region where QML provides the best improvements over the Pseudo-$\Cl$
estimator, and so could provide sufficiently tighter constraints when applied to
these data-sets.

Since QML methods deal with data at the pixel-level, they are well suited for
dealing with contaminants and effects that can only be described accurately in
terms of pixels in the maps. One such problem is the effect of spatially varying
noise over survey area, which could arise from different seeing conditions
encountered as a telescope surveys the sky or the properties of the detector
evolving as data is taken. These effects can easily be incorporated into the QML
estimator through an appropriate modification of the noise matrix
$\mathbfss{N}$, whereas the Pseudo-$\Cl$ method utilises Fourier transforms and
these pixel-level effects get diffused over a wide $\ell$ range and thus become
harder to model. This could further reduce the optimality of the Pseudo-$\Cl$
estimator. We leave a dedicated investigation of how such effects affect
the two estimators to future work.

To conclude, we have shown that the Pseudo-$\Cl$ estimator is close to optimal
on small scales for a simplified \textit{Euclid}-like weak lensing survey.
Despite this, the QML estimator is better suited for a variety of applications,
including $E$/$B$-mode separation, complex noise patterns, and complicated
survey geometries. With systematics expected to dominate the error budget of
upcoming surveys it is increasingly important to demonstrate the consistency of
results derived from different analysis pipelines - the fast, publicly available
implementation of the QML estimator that we have presented in this work
represents a significant step forward in this regard.

\section*{Data availability}

All data presented in this work has been generated by the authors. The code to
do so can be found on our GitHub repository located at
\href{https://github.com/AlexMaraio/WeakLensingQML}{\texttt{https://github.com/AlexMaraio/WeakLensingQML}
\faicon{github}}.

\section*{Acknowledgements}

AM would like to thank all members of Lensing Coffee at the IfA for many useful
conversations and invaluable support. AH thanks Uro\v s Seljak for useful
discussion. AH and AT are supported by a Science and Technology Facilities
Council (STFC) Consolidated Grant. For the purpose of open access, the author
has applied a Creative Commons Attribution (CC BY) licence to any Author
Accepted Manuscript version arising from this submission.

\bibliographystyle{mnras}
\bibliography{References/references}

\appendix

\section{Ratio of numeric to analytic Fisher}
\label{app:Full_cl_fisher_ratio}

\begin{figure*}
    \centering
    \includegraphics[width=0.85\textwidth]{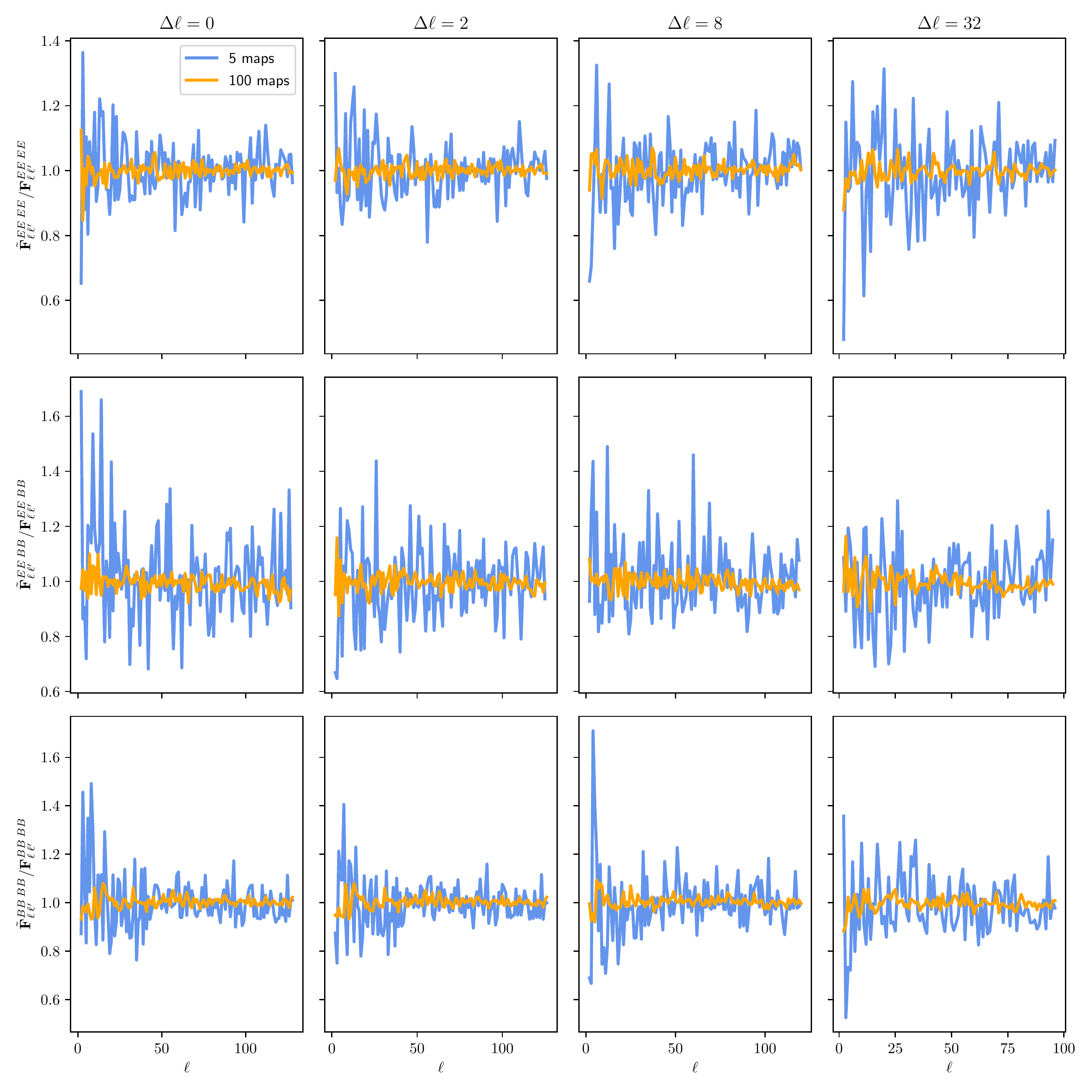}
    \caption{Ratio of our numerically-derived $\Cl$-Fisher matrix to the
    analytic result for a map resolution of $\Nside = 64$, presented for the
    cases where we average over five and one hundred maps. Here, we plot the
    $EE$-$EE$, $EE$-$BB$, and $BB$-$BB$ components separately, with varying
    off-sets from the diagonal in the different columns. We see good agreement
    between our estimator and existing results for all combination of spectra
    and off-sets, and so deduce that our numerical estimate is consistent
    with the analytic result.}
    \label{fig:Numeric_to_analytic_Fisher_ratio_grid}
\end{figure*}

Figure \ref{fig:Numeric_to_analytic_Fisher_ratio_grid} shows the ratio of our
numerically computed $\Cl$ Fisher matrix to that computed using analytic methods
for a number of different off-set values from the diagonal. Here, we see that
all curves simply exhibit random scatter around unity which shows that our
numerical estimates of the Fisher matrix is an unbiased estimate of the true
values. The residual noise in the Fisher matrix gives rise to negligible 
differences in parameter confidence contours.

\section{Sensitivity to apodisation}
\label{app:Apodising_mask}

\begin{figure}
    \includegraphics[width=\columnwidth]{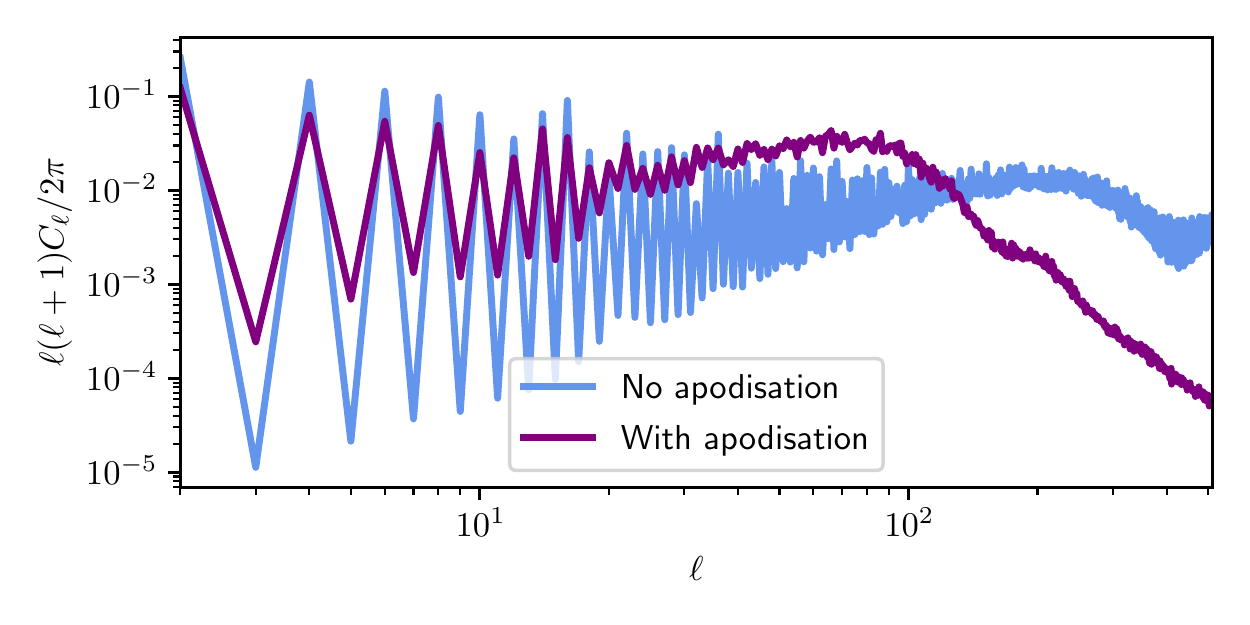}
    \vspace*{-0.5cm}
    \caption{Power spectrum of our mask, with stars included, for cases with and
      without apodisation applied.}
    \label{fig:Mask_power_spectrum_apodisation}
\end{figure}

\begin{figure}
    \includegraphics[width=\columnwidth]{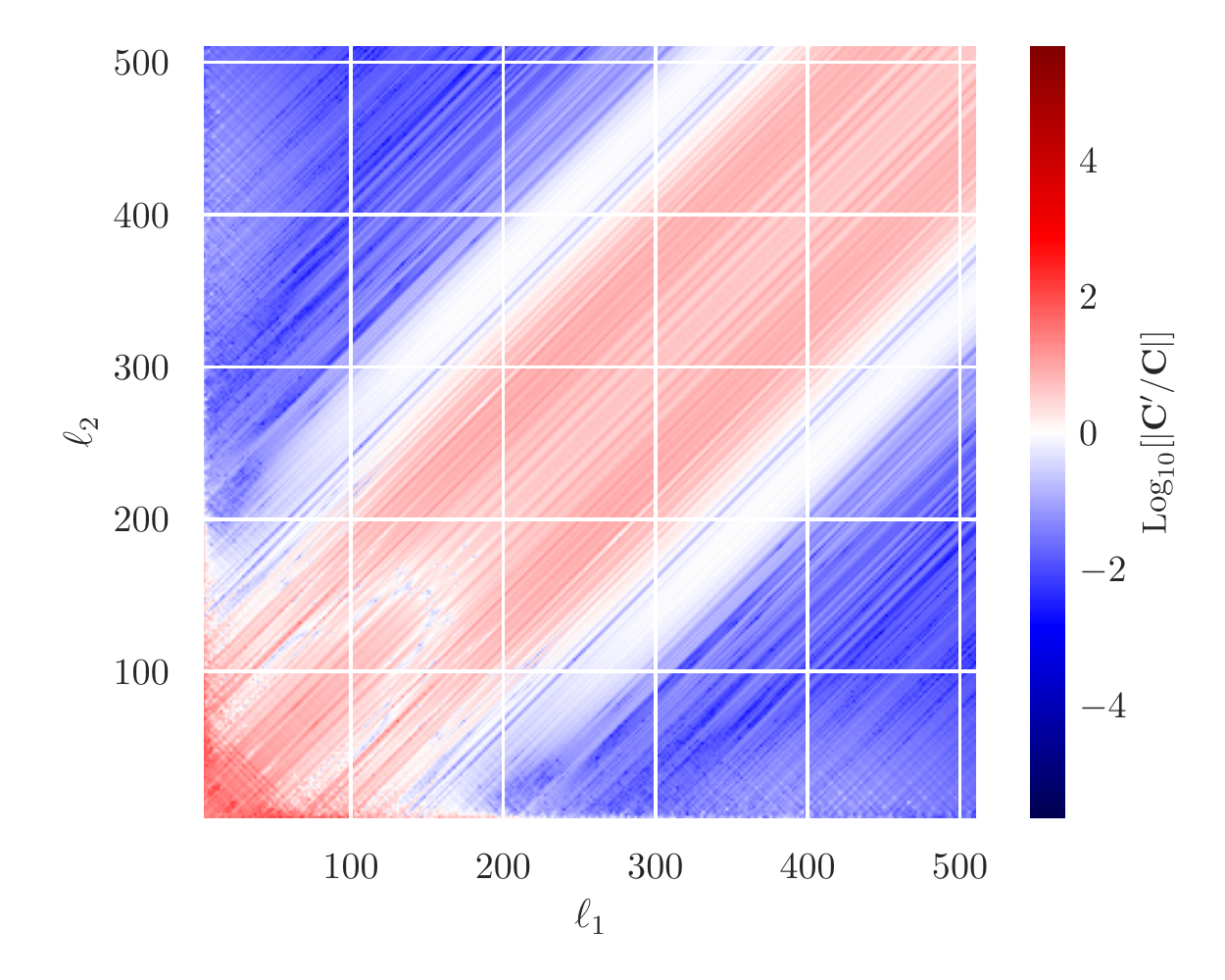}
    \vspace*{-0.5cm}
    \caption{Ratio of the analytic Pseudo-$\Cl$ covariance matrix for the
      $\ClEE$ power spectrum for the cases with ($\mathbfss{C}^{\prime}$) and
      without ($\mathbfss{C}$) mask apodisation applied.}
    \label{fig:PCl_cov_apo_ratio}
\end{figure}

Previously, we have discussed how apodisation of the mask is not required for
QML methods whereas there are certain advantages to doing so for the
Pseudo-$\Cl$ method. This is because apodisation reduces the effects of sharp
edges that may be present in the mask. To investigate the effects of
apodisation, we have applied a $2^{\circ}$ apodisation using the
$\mathcal{C}^{2}$ scheme as described in~\cite{Alonso:2018jzx} to our mask,
including stars. We note that for this apodisation scale and scheme, the sky
area reduces from $\fsky = 33 \, \%$ to $\fsky = 22 \, \%$.
Figure~\ref{fig:Mask_power_spectrum_apodisation} shows the power spectrum of our
mask with and without apodisation applied. Here, we see that the effect of
apodisation is to vastly reduce the small-scale power of the mask. The effect of
this suppression of small-scale power results in the reduction in long-range
correlations in the covariance matrix, as can be shown from
Equation~\ref{eqn:exact_pcl_covariance}, and thus the computation and inversion
of the mixing matrix should be more accurate when apodisation is applied. The
ratio of the analytic Pseudo-$\Cl$ covariance matrix for the $\ClEE$ spectrum
for the cases of with and without apodisation is shown in
Figure~\ref{fig:PCl_cov_apo_ratio}. This shows that for values along and close to
the diagonal, the loss of sky area causes significant increases in the variances
in the power spectrum. For mode-pairs that are highly separated we see
a notable decrease in their covariances, which once again can be seen from
Equation~\ref{eqn:exact_pcl_covariance}.

\begin{figure}
    \includegraphics[width=\columnwidth]{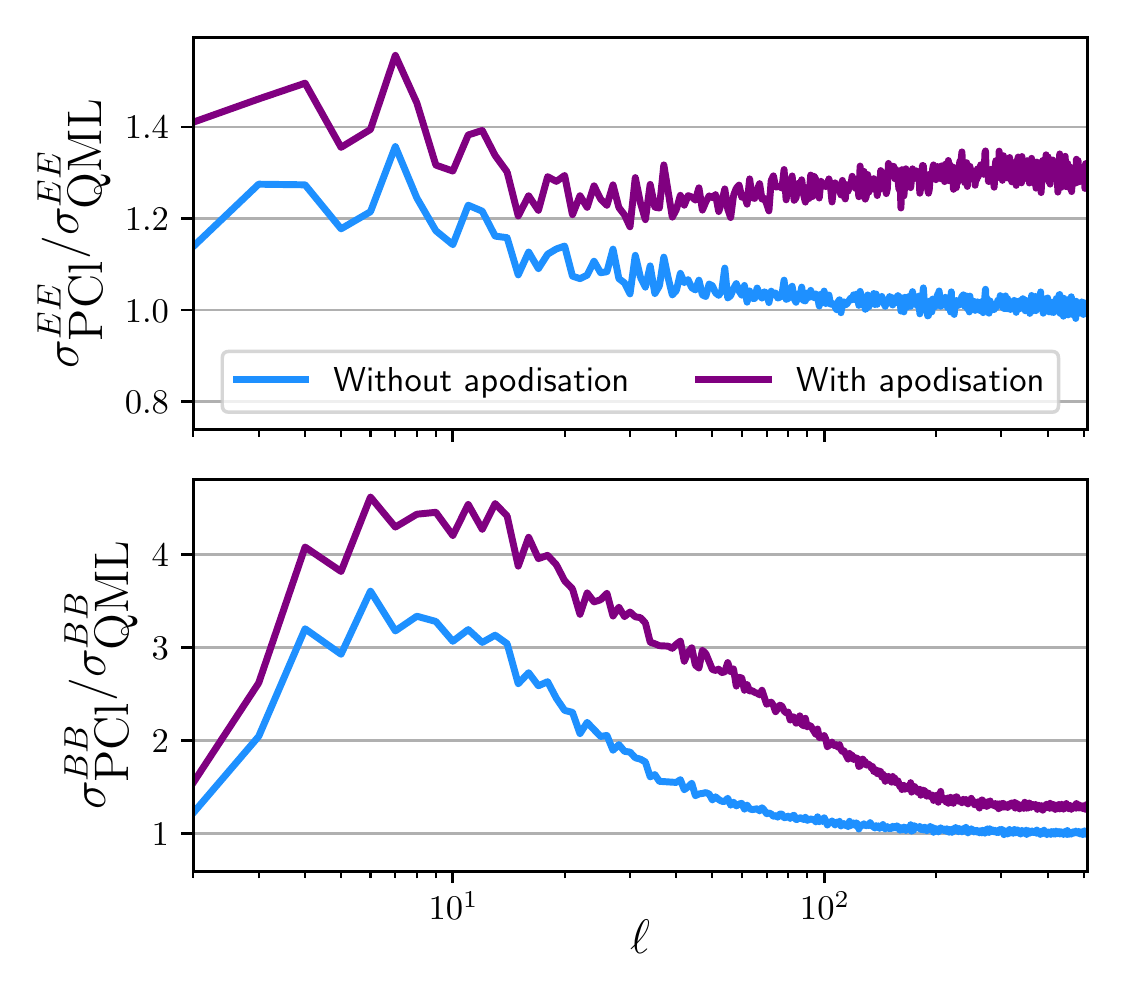}
    \caption{Ratio of $\Cl$ errors of the Pseudo-Cl method with respect to
    our QML estimator for the cases with and without apodisation.}
    \label{fig:Cl_err_ratio_with_apo}
\end{figure}

Figure~\ref{fig:Cl_err_ratio_with_apo} shows the ratio of the errors for the
Pseudo-$\Cl$ (the square-root of the diagonal of the covariance matrix) with
respect to our QML estimator for the case of with and without apodisation. Here,
we see that the effect of apodisation is to increase the errors of the
Pseudo-$\Cl$ method - which is a direct result in the loss of sky area that
apodisation produces.

\begin{figure}
    \includegraphics[width=\columnwidth, trim={0cm 0cm 0cm 0cm}, clip]{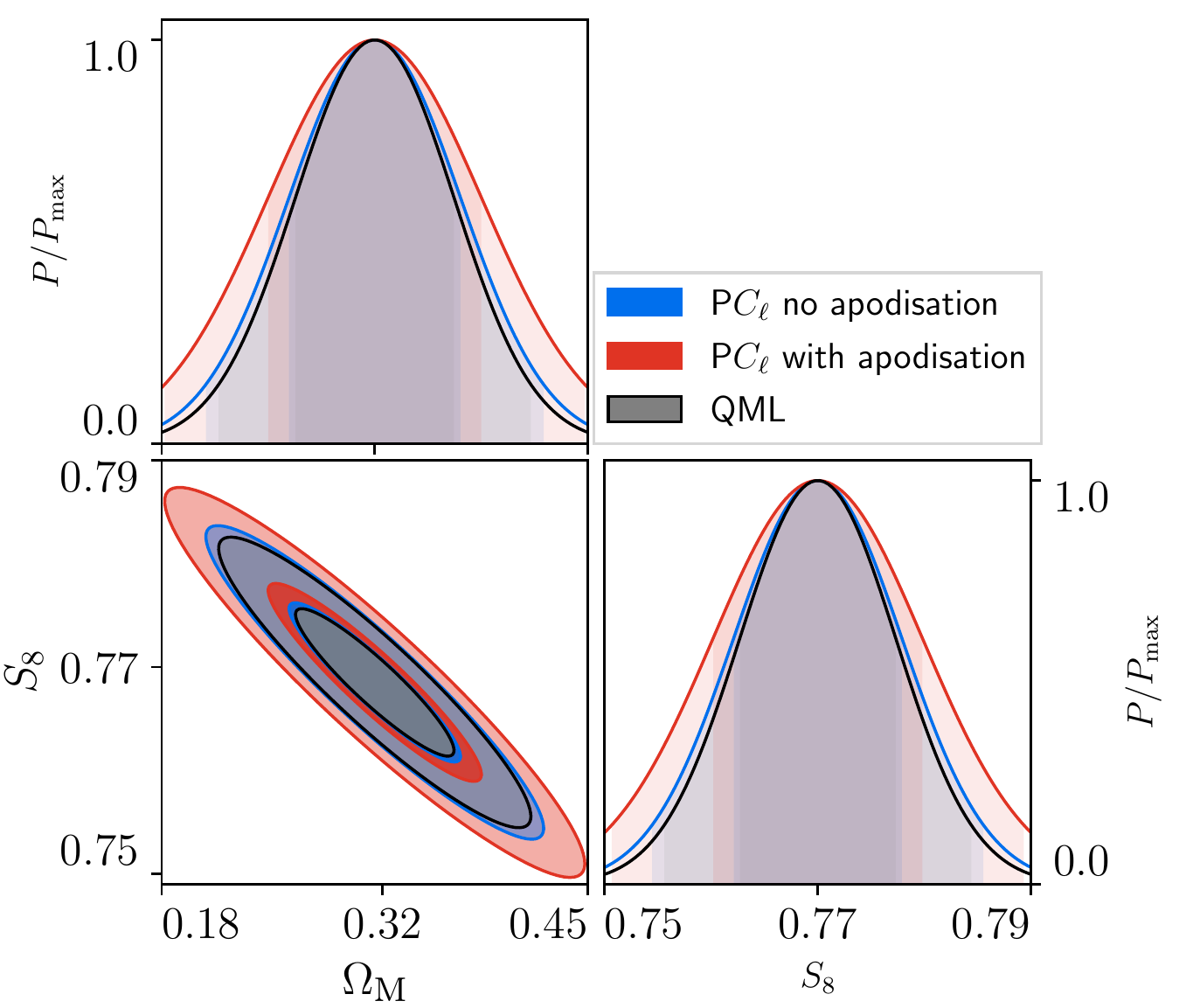}
    \vspace*{-0.5cm}
    \caption{Fisher parameter constraints comparison between QML and
      Pseudo-$\Cl$ where both estimators have a maximum multipole of $\lmax =
      512$ and for the case where apodisation has been applied for the
      Pseudo-$\Cl$ method. We see a large broadening for the Pseudo-$\Cl$
      contour with apodisation applied, which is consistent with the loss of
      sky area that apodisation results in.}
    \label{fig:Parameter_constraints_apodisation}
\end{figure}

The covariance matrix for the case where we have applied apodisation can then be
propagated into parameter constraints, which is shown in
Figure~\ref{fig:Parameter_constraints_apodisation}. Here, we see that the direct
loss of sky area associated with apodisation results in broadened parameter
contours which is not offset by the decrease in long-range correlations that
apodisation suppresses in the covariance matrix.

\bsp	
\label{lastpage}
\end{document}